\documentclass{lmcs}
\usepackage[utf8]{inputenc}
\pdfoutput=1

\usepackage{lastpage}
\lmcsdoi{18}{3}{7}
\lmcsheading{}{\pageref{LastPage}}{}{}%
{Dec.~01,~2020}{Jul.~28,~2022}{}

\title{Modularising Verification of Durable Opacity} 

\author[E.~Bila]{Eleni Bila}[a]	
\address{University of Surrey}	
\email{e.vafeiadibila@surrey.ac.uk }  
\email{b.dongol@surrey.ac.uk }  
\newlength{\myrowsepc}

\author[J.~Derrick]{John Derrick}[b]	
\address{University of Sheffield}
\email{j.derrick@sheffield.ac.uk }  
\email{s.doherty@sheffield.ac.uk }  
\thanks{Derrick and Doherty are supported by EPSRC project EP/R032351/1. Dongol is supported by EPSRC Grants EP/R032556/1, EP/V038915/1 and EP/R025134/2. Bila and Dongol are supported by VeTSS project ``Persistent Safety and Security''. Wehrheim is partially supported by DFG grant \mbox{WE2290/12-1}}	

\author[S.~Doherty]{Simon Doherty}[b]	

\author[B.~Dongol]{Brijesh Dongol}[a]	

\author[G.~Schellhorn]{Gerhard Schellhorn}[c]	
\address{University of Augsburg}	
\email{schellhorn@informatik.uni-augsburg.de}  

\author[H.~Wehrheim]{Heike Wehrheim}[d]	
\address{University of Oldenburg}	
\email{heike.wehrheim@uol.de}  

\usepackage{amssymb}
\usepackage{verbatim}

\usepackage{graphicx}
\usepackage{listings}
\usepackage{xcolor}
\usepackage{algorithmic}
\usepackage{amsmath,amsthm}
\usepackage{listings}
\backuprestoremacros{proof,endproof,qed,QED,Qed}{\usepackage{oz}}
\usepackage{xspace}
\usepackage{comment}
\usepackage{pifont}


\usepackage{tikz}

\usetikzlibrary{positioning,shapes,arrows,calc,decorations.markings,math,arrows.meta}
\usepackage{hyperref}
\theoremstyle{plain}
\newtheorem{theorem}[thm]{Theorem}
\newtheorem{lemma}[thm]{Lemma}

\newtheorem{definition}[thm]{Definition}
\newtheorem{corollary}[thm]{Corollary}

\newcommand*\circled[1]{\tikz[baseline=(char.base)]{
            \node[shape=circle,draw,inner sep=1pt] (char) {#1};}}

\newcommand{\DTMS}{\text{\sc dTMS2}\xspace}
\newcommand{\TMS}{\text{\sc TMS2}\xspace}

\newcommand{\NOREC}{\text{\sc NOrec}\xspace}
\newcommand{\DNOREC}{\text{\sc dNOrec}\xspace}
\newcommand{\DLNOREC}[1]{\text{\sc dNOrec[\ensuremath{#1}]}\xspace}
\newcommand{\LB}[1]{\ensuremath{B[#1]}\xspace}

\newcommand{\AM}{\text{\sc AM}\xspace}
\newcommand{\CM}{\text{\sc CM}\xspace}

\newcommand{\IMPL}{\text{\sc IMPL}\xspace}
\newcommand{\DIMPL}{\text{\sc dIMPL}\xspace}
\newcommand{\DLIMPL}[1]{\text{\sc dIMPL[\ensuremath{#1}]}\xspace}

\newcommand{\ops}{{\it ops}}
\newcommand{\Ops}{\mathit{\Sigma}}
\newcommand{\Tids}{T}
\newcommand{\ev}{\mathit{e}}

\newcommand{\resp}{{\it res}}
\newcommand{\crash}{\mathit{c}}

\newcommand{\Loc}{\mathit{Loc}}
\newcommand{\Val}{{\it Val}}

\newcommand{\abort}{\texttt{abort}}

\newcommand{\tmbegin}{\texttt{TMBegin}}
\newcommand{\tmbegino}{\texttt{TMBegin}({\tt ok})}
\newcommand{\tmbegina}{\texttt{TMBegin}({\tt abort})}
\newcommand{\tmend}{\texttt{TMCommit}}
\newcommand{\tmendc}{\texttt{TMCommit}({\tt commit})}
\newcommand{\tmenda}{\texttt{TMCommit}({\tt abort})}
\newcommand{\tmread}{\texttt{TMRead}}

\newcommand{\tmreada}{\texttt{TMRead}({\tt abort})}
\newcommand{\tmwrite}{\texttt{TMWrite}}
\newcommand{\tmwriteo}{\texttt{TMWrite}({\tt ok})}
\newcommand{\tmwritea}{\texttt{TMWrite}({\tt abort})}

\newcommand{\readob}{\texttt{LibRead}}
\newcommand{\writeob}{\texttt{LibWriteSet}}
\newcommand{\acquireob}{\texttt{LibAcquire}}
\newcommand{\releaseob}{\texttt{LibRelease}}

\newcommand{\reffig}[1]{Figure~\ref{#1}}
\newcommand{\refthm}[1]{Theorem~\ref{#1}}

\newcommand{\refsec}[1]{\autoref{#1}}

\newcommand{\refdef}[1]{Definition~\ref{#1}}

\newcommand{\beginInv}[1]{\ensuremath{inv_{#1}(\tmbegin)}}
\newcommand{\beginResp}[1]{\ensuremath{res_{#1}(\tmbegino)}}
\newcommand{\readInv}[2]{\ensuremath{inv_{#1}(\tmread(#2))}}
\newcommand{\readResp}[2]{\ensuremath{res_{#1}(\tmread(#2))}}
\newcommand{\writeInv}[2]{\ensuremath{inv_{#1}(\tmwrite(#2))}}
\newcommand{\writeResp}[1]{\ensuremath{res_{#1}(\tmwriteo)}}
\newcommand{\commitInv}[1]{\ensuremath{inv_{#1}(\tmend)}}
\newcommand{\commitResp}[1]{\ensuremath{res_{#1}(\tmendc)}}

\newcommand{\abortResp}[1]{\ensuremath{res_{#1}(op(\abort))}}
\newcommand{\run}[1]{run_t}

\newcommand{\sif}{\texttt{if}}
\newcommand{\sthen}{\texttt{then}}
\newcommand{\selse}{\texttt{else}}

\newcommand{\doCommitReadOnly}[2]{\ensuremath{\texttt{DoCommitReadOnly}_{#1}(#2)}}
\newcommand{\doCommitWriter}[1]{\ensuremath{\texttt{DoCommitWriter}_{#1}}}
\newcommand{\doRead}[2]{\ensuremath{\texttt{DoRead}_{#1}(#2)}}
\newcommand{\doWrite}[2]{\ensuremath{\texttt{DoWrite}_{#1}(#2)}}

\newcommand{\doLibRead}[2]{\ensuremath{\texttt{DoLibRead}_{#1}(#2)}}

\newcommand{\OreadInv}[2]{\ensuremath{inv_{#1}(\readob(#2))}}
\newcommand{\OreadResp}[2]{\ensuremath{res_{#1}(\readob(#2))}}
\newcommand{\OwriteInv}[2]{\ensuremath{inv_{#1}(\writeob(#2))}}
\newcommand{\OwriteResp}[2]{\ensuremath{res_{#1}(\writeob(#2))}}

\newcommand{\wone}[1]{\ensuremath{\texttt{W1}_{#1}}}
\newcommand{\wtwo}[1]{\ensuremath{\texttt{W2}_{#1}}}
\newcommand{\wthree}[1]{\ensuremath{\texttt{W3}_{#1}}}
\newcommand{\wfour}[1]{\ensuremath{\texttt{W4}_{#1}}}
\newcommand{\wfive}[1]{\ensuremath{\texttt{W5}_{#1}}}
\newcommand{\wsix}[1]{\ensuremath{\texttt{W6}_{#1}}}
\newcommand{\wseven}[1]{\ensuremath{\texttt{W7}_{#1}}}
\newcommand{\weight}[1]{\ensuremath{\texttt{W8}_{#1}}}

\newcommand{\rone}[1]{\ensuremath{\texttt{R1}_{#1}}}

\newcommand{\action}[3]{\ensuremath{
\begin{array}[t]{ll}
\multicolumn{2}{l}{#1}\\
\textsf{Pre: }&#2\\
\textsf{Eff: }&#3
\end{array}
}}

\newcommand{\pcNotStarted}{\statusText{notStarted}}
\newcommand{\pcBeginPending}{\statusText{beginPending}}

\newcommand{\pcReady}{\statusText{ready}}
\newcommand{\pcDoWrite}{\statusText{doWrite}}
\newcommand{\pcWriteResp}{\statusText{resWrite}}
\newcommand{\pcDoRead}{\statusText{doRead}}
\newcommand{\pcReadResp}{\statusText{resRead}}

\newcommand{\pcDoLibRead}{\statusText{doLibRead}}
\newcommand{\pcReadLibResp}{\statusText{resLibRead}}
\newcommand{\pcDoLibWrite}{\statusText{doLibWriteSet}}
\newcommand{\pcWriteLibResp}{\statusText{resLibWriteSet}}

\newcommand{\pcDoCommit}{\statusText{doCommit}}
\newcommand{\pcCommitResp}{\statusText{resCommit}}
\newcommand{\pcCancelPending}{\statusText{cancelPending}}
\newcommand{\pcCommitted}{\statusText{committed}}
\newcommand{\pcAborted}{\statusText{aborted}}
\newcommand{\pcCrashed}{\statusText{crashed}}

\newcommand{\pcwone}{\statusText{W1}}
\newcommand{\pcwtwo}{\statusText{W2}}
\newcommand{\pcwthree}{\statusText{W3}}
\newcommand{\pcwfour}{\statusText{W4}}
\newcommand{\pcwfive}{\statusText{W5}}
\newcommand{\pcwsix}{\statusText{W6}}
\newcommand{\pcwseven}{\statusText{W7}}
\newcommand{\pcweight}{\statusText{W8}}

\usepackage{tikz}
\usetikzlibrary{positioning}
\newcommand{\firststate}{\mathtt{first}}
\newcommand{\repl}[2]{\ensuremath{\mathtt{repl}(#1,#2)}}
\newcommand{\replfun}{\ensuremath\mathtt{repl}}
\newcommand{\exec}{\mathtt{exec}}
\newcommand{\trace}{\mathtt{trace}}

\newcommand{\statusText}[1]{\text{#1}}
\newcommand{\bbL}{\mathbb{L}}
\newcommand{\bbS}{\mathbb{S}}
\newcommand{\duraut}[1]{\textsc{DurAut}(#1)}
\newcommand{\durlin}[1]{\textsc{DurLin}(#1)}

\begin{document}

\begin{abstract}
Non-volatile memory (NVM), also known as persistent memory, is an emerging paradigm for memory that preserves its contents even after power loss. 
NVM is widely expected to become ubiquitous, and hardware architectures are already providing support for NVM programming. 
This has stimulated interest in the design of novel concepts ensuring correctness of concurrent programming abstractions in the face of persistency and in the development of associated verification approaches. 

Software transactional memory (STM) is a key programming abstraction that supports concurrent access to shared state. In a fashion similar to linearizability as the correctness condition for concurrent data structures, there is an established notion of correctness for STMs known as opacity. 
We have recently proposed {\em durable opacity} as the natural extension of opacity to a setting with 
non-volatile memory. Together with this novel correctness condition, we designed a verification technique based on refinement. In this paper, we extend this work in two directions.
First, we develop a durably opaque version of NOrec (no ownership records), an existing STM algorithm proven to be opaque. 
 Second, we modularise our existing verification approach by separating the proof of durability 
 of memory accesses from the proof of opacity. For NOrec, this allows us to {\em re-use} an 
 existing opacity proof  and complement it with a proof of the durability of accesses to shared state.

\end{abstract}

\maketitle

\section{Introduction}
Non-volatile memory (NVM) promises the combination of the density and non-volatility of NAND Flash-based solid-state disks (SSDs) with the performance of volatile memory (RAM). The term {\em persistent memory} is used to describe an NVM technology that presents two characteristics: {\bf (1)} directly byte-addressable access from the user space by using byte-addressable operations and {\bf (2)} preservation of its contents even after system crashes and power failures.
NVM is intended to be used as an intermediate layer between traditional
volatile memory (VM) and secondary storage, and has the potential to vastly
improve system speed and stability.
Speed-ups of 2-3 orders of magnitude are likely to be feasible over and above hard disks.
Furthermore, software that uses NVM has the potential to be more robust; in case of a crash, a system state before the crash may be recovered using contents from NVM, as opposed to being restarted from secondary storage. 
For these reasons alone, NVM is widely expected to become ubiquitous, and hardware architectures are already providing support for NVM programming. 

However, writing correct NVM programs is extremely difficult, as the semantics of persistency can be unclear.
Furthermore, because the same data is stored in both a volatile and non-volatile manner, and because NVM is updated at a slower rate than VM, recovery to a consistent state may not always be possible. This is particularly true for concurrent systems, where coping with NVM requires introduction of additional synchronisation instructions into a program.
Such instructions are already supported by Intel-x86 and ARMv8.

This has led to work on the design of the first persistent concurrent programming
abstractions, so far mainly concurrent data structures~\cite{DBLP:journals/pacmpl/ZurielFSCP19,DBLP:conf/ppopp/FriedmanHMP18,DBLP:conf/pldi/FriedmanPR21,volos2011mnemosyne,DBLP:conf/fast/VenkataramanTRC11}. 
To support the reasoning about correctness for these abstractions working over NVM, a coherent notion of correctness is needed. 
Such a notion for concurrent data structures has been  defined by Izraelevitz et al.~\cite{DBLP:conf/wdag/IzraelevitzMS16} (known as {\em durable linearizability}) which naturally  generalises the standard linearizability correctness condition~\cite{HeWi90}.
A first proof technique for showing durable linearizability   has been proposed by 
Derrick et al.~\cite{DBLP:conf/fm/DerrickDDSW19}.

In this paper we investigate another key programming abstraction known as {\em Software Transactional Memory} (STM) that supports concurrent access to shared state. 
STM is a mechanism that provides an illusion of atomicity in concurrent programs and aims to reduce the burden on programmers of implementing complicated synchronisation
mechanisms. 
The analogy of
STM is with database transactions, which perform a series of accesses/updates to shared
data (via read and write operations) atomically in an all-or-nothing manner. Similarly
with an STM, if a transaction commits, all its operations succeed, and in the aborting
case, all its operations fail. 
STMs are now part of mainstream programming, e.g., the
ScalaSTM library, a new language feature in Clojure that uses an STM implementation internally for all data manipulation and the G++ 4.7 compiler (which supports STM features directly in the compiler).

In a fashion similar to linearizability as the correctness condition for concurrent data structures, there is an established notion of correctness for STMs known as {\em opacity}~\cite{GuerraouiK08}.  Overall,  opacity guarantees that committed transactions  
appear as if they are executed atomically, at some unique point in time,
and aborted transactions, as if they did not execute at all.  Amongst
other things, opacity also guarantees that all reads that a
transaction performs are valid with respect to a single memory
snapshot.


A fundamental challenge when developing STMs for persistent memory is to ensure a correct recovery after a crash.
This requires that, at any point in the execution of the program,
the persistent state must be sufficient to enable the recovery procedure to recreate
an appropriate consistent state. Verification of STMs has to show that this is achieved by 
the proposed algorithm, i.e., that enough data is persisted and the recovery procedure correctly uses this data to guarantee opacity.

In this paper, we investigate STMs and their correctness via opacity on non-volatile memory architectures. 
Doing this entails a number of steps. First, the correctness criterion
of opacity has to be adapted to cope with crashes in system
executions. Second, STM algorithms have to be extended to deal with
the coexistence of volatile and non-volatile memory during execution
and need to be equipped with recovery operations. Third, proof
techniques for opacity need to be re-investigated to make them usable
for durable opacity. 

In our prior work~\cite{DBLP:conf/forte/BilaDDDSW20},
we have addressed the steps above as follows. The first step is addressed by defining a notion of correctness called {\em durable
opacity}, which generalises opacity in the same
way that durable linearizability~\cite{DBLP:conf/wdag/IzraelevitzMS16} generalises linearizability for NVM architectures. Durable opacity requires
executions of STMs to be opaque even if they are interspersed with
crashes.  The second step is addressed by {\em developing}
a durable version of the Transactional Mutex
Lock~\cite{DalessandroDSSS10}. Finally, the third step is addressed by {\em proving} durable
opacity of this new algorithm using a refinement-based approach.

This paper extends prior work~\cite{DBLP:conf/forte/BilaDDDSW20} via
the development of a {\em modular approach} to verifying durable opacity.
Our new approach is inspired by the modularised verification 
of a filesystem for flash memory
\cite{DBLP:conf/ifm/PfahlerEBSR17,Bodenmueller-fac2022}.
The proof technique separates the
proof of opacity (perceived atomicity of transactions)
from the proof of durability (correct handling of non-volatile memory). Our proof technique assumes
the existence of an STM that has been verified to be opaque by proving that it refines the
specification TMS2~\cite{DGLM13} (which itself has been shown to satisfy opacity~\cite{Lesani2012}).  
This refinement proof is then {\em re-used} to construct a durably opaque version of the STM. We exemplify our technique by extending the No-Ownership-Records
(\NOREC) STM of Dalessandro et al.~\cite{DBLP:conf/ppopp/DalessandroSS10}
to create a durable \NOREC.

\begin{figure}\label{fig:TMS2-DTMS2}
\begin{tabular}{p{0.49\textwidth} p{0.49\textwidth}}
  \vspace{0pt} \centering\includegraphics[scale=0.4]{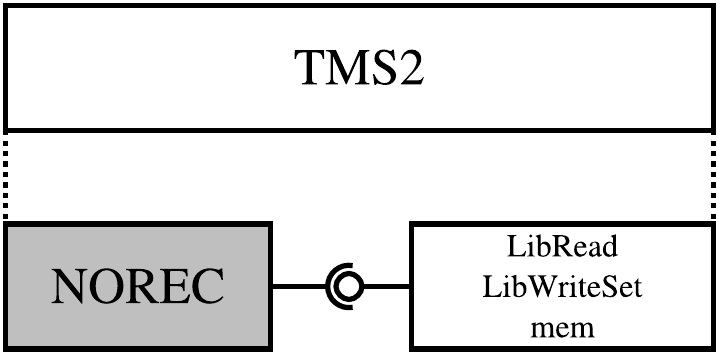} &
  \vspace{0pt} \centering\includegraphics[scale=0.4]{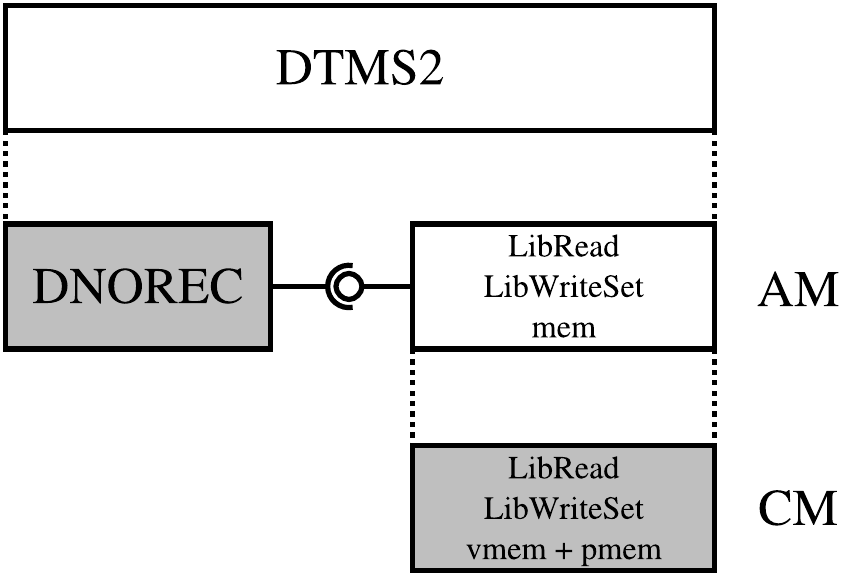}
  \end{tabular}
\caption{Original proof of opacity (left) vs. proof of durable opacity (right)}
 
\end{figure}


\reffig{fig:TMS2-DTMS2} illustrates our approach. The original \NOREC algorithm
is shown to the left. It has already been proven opaque by showing that it refines
(dashed lines) the TMS2 automaton by Lesani et al.~\cite{Lesani2012} using the PVS prover.
The algorithm can be thought of as having an implicit interface
to main memory {\em mem}  (indicated by the \includegraphics[scale=0.2]{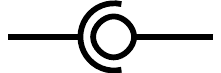} symbol),
which allows to read and write memory cells. Since \NOREC is a lazy algorithm,
writing is confined to committing a write set to ensure that two transactions cannot commit
their write set at the same time,  producing a mixed result that 
would contradict opacity. \NOREC enforces that there is at most one transaction
committing a write set any time. Our approach will first make this interface
with operations {\tt LibRead} and {\tt LibWriteSet} explicit  and call it \AM.
The tricky bit in defining the interface is the enforcement of  the constraint of a single committer as
an ownership annotation for \AM \footnote{Other algorithms, like  TL2 enforce disjoint write sets to ensure that there are no conflicts, which would
result in a similar interface with a modified concept of ownership (in this case about memory locations).}. (This annotation parallels the use of an auxiliary variable in the original PVS proof of \NOREC~\cite{Lesani2012}.)

It can then be observed that 
 if a) all reads and writes to memory were directly to persistent
memory and b) committing a write set is atomic,  then the resulting
algorithm is already durable opaque since the content of memory is preserved on a crash.  Crashes in the middle of commits that could lead to a state that is not compatible with durable opacity are then avoided.
As a consequence, we can reuse the original opacity proof with only minor adjustments. The main change is that using
the abstract \DTMS automaton (\reffig{fig:TMS2-DTMS2}, right)
to express durable opacity  adds the proof obligation
that a crash does indeed not have any relevant effect.
Since the original opacity proof is by far the most complex proof needed,
reusing it saves a lot of work compared to verification from scratch. Of course,  assumptions a) and b) above are not realistic assumptions
when viewing \AM as an \emph{implementation}. However, \AM can also be viewed
as a \emph{specification} of a library  that can be refined  to a non-atomic, concurrent implementation.
We define such an implementation \CM. It basically uses volatile memory {\em vmem} as a cache for persistent memory {\em pmem}.
A logging mechanism ensures that a recovery procedure that runs on restarting from a crash can undo the effects of a partially completed transaction. The correctness proof for the refinement
then is completely {\em separate} from the main proof. It shows that  \CM is  a durable linearizable
implementation of  \AM. We then prove in general, that two refinements constructed in this way together
always give a proof of durable opacity for an algorithm which combines the two implementations shown
in grey in the figure (written \DLNOREC{\CM}  for our case).

The approach of this paper therefore can be viewed as a blueprint for a modular strategy,
that allows to transform an STM implementation that is opaque to a durably opaque one.
In particular,  we believe our modularisation technique can be used on any transactional
memory algorithm that uses a write-log and serialises commits
\cite{DBLP:conf/ppopp/DalessandroSS10, ShavTL2, RingSTM}.

The difficult bit for each algorithm will be the definition of an interface \AM with suitable
ownership conditions, that ensure that its implemenation \CM only has to deal with a
suitably restricted form of concurrency (here: no two commits at the same time).
However such restrictions must have already been relevant for proving opacity
of the original algorithm, so similar to our case it should be possible to move them
to constraints on \AM,  to reuse the original proof and to construct a separate refinement
to \CM.

We  mechanise the proof of durable linearizability of the library
in the theorem prover KIV~\cite{KIV-Haehnle2022}. KIV is also used to mechanise a general result
on using refinement in a context which specialises to the result that the two refinement proofs together
imply durable opacity of the final algorithm with library calls.
These proofs are available online
\cite{KIV-DNOREC}.

\subsection*{Overview} This paper is organised as follows. In \refsec{sec:foundations}, we give background for this paper, and present the execution model and the formal definitions of (durable) linearizability and (durable) opacity. In \refsec{sec:prov-opac-meth}, we present the use of IOA to verify correctness of durable concurrent objects.  Our modular proof technique is described in \refsec{sec:modular-proofs}, which describes the transformation of an opaque algorithm to satisfy durable opacity, the modularisation of memory accesses using an abstract library and its fine-grained refinement of the abstract library to a concrete library. Both the modularisation and library refinement steps are guaranteed to preserve durable opacity. We cover related work in \refsec{sec:related-work}. 

\section{Foundations}
\label{sec:foundations}

We start by explaining some basic assumptions we make about the memory model and 
by explaining how persistent and volatile memory interact. We then define the
correctness conditions relevant for our approach.  These  are {\em linearizability}~\cite{HeWi90} and {\em opacity}~\cite{GuerraouiK08}. Linearizability (or better to say, its adaption for NVM) is part of our proof method, and the NVM-version of opacity, durable opacity, is the concurrent correctness criterion we intend to prove for STMs. 
Both correctness conditions 
formalise some form of atomicity in which a block of code executes seemingly
atomically in an all-or-nothing manner. The difference lays in
the level of atomicity: for linearizability blocks of code describe
one operation of a concurrent data structure; for opacity we also have
blocks of code for specific operations, and in addition group such
operations into {\em transactions}.

\subsection{Memory model, crashes and recovery} \label{sec:mem-model}

We assume that the shared state consists of a set $\Loc$ of locations
and contains values from a set $\Val$. Threads can concurrently access
locations, and we assume these accesses to be sequentially consistent
(SC~\cite{Lamport79}).

Algorithms running on NVM architectures operate on two versions of
memory: {\em persistent} and {\em volatile} memory (later denoted as
$pmem$ and $vmem$, respectively). In an NVM architecture, a write to
some location $l \in \Loc$ first of all only modifies
$vmem(l)$. Volatile memory is then occasionally {\em flushed} to
persistent memory by the system. This updates the value of persistent
memory to the value currently in volatile memory for 
location $l$. The programmer can also enforce such a flush to location
$l$ by executing \texttt{flush(l)}\footnote{We use typewriter font to refer to program code.}, which is modelled by an update
that sets $pmem(l)$ to $vmem(l)$.

When a crash occurs, the contents of volatile memory is lost and that
of persistent memory is kept. We assume that immediately after a crash
$vmem$ is (re)set to $pmem$, thus any writes to $vmem$ that have not
been flushed will be lost.

The implementations of concurrent data structures or STM algorithms have to ensure that shared memory is
kept in a consistent state, despite these losses. To this end, they
need to persist enough data (i.e., flush it) to be able to bring
shared memory back to a consistent state after crashes. For our implementations, we assume that such a {\em
  recovery} step is automatically executed by the algorithms after every crash. In
our models of the algorithms, we formalise this by a single atomic operation ${\mathit
  crashRecovery}$. Note that this is not a strict requirement of durable opacity, i.e., durable opacity (like durable linearizability) admits other algorithms in which the crash and recovery occur as two separate steps.

The execution with persistent and volatile memory applies to actual
{\em implementations}, i.e., the low-level descriptions of STM
algorithms with all the implementation details filled
in. Implementations are one conceptual entity within our reasoning
technique based on {\em
  refinement}~\cite{DBLP:books/daglib/0032291}. Refinement compares
abstract {\em specifications} to concrete implementations. The purpose
of an abstract specification is to fix the allowed execution
traces. Abstract specifications are hence not subject to specific
forms of execution with volatile and persistent memory; they are
allowed to (and should) abstract from implementation details. Thus, we often develop intermediate models that interact directly to NVM (bypassing volatile memory), with more realistic interactions between volatile and persistent memory only appearing in the final
implementation (see~\reffig{fig:CDT}). 

\subsection{Histories}
Both correctness conditions are formalised in terms of a  {\em
  history}, which is a sequence of {\em events}.  An event is either
(1) an invocation ({\em inv}) or (2) a response ({\em res}) of an
operation $op$ out of a set of operations $\Ops$ or (3) a system-wide
crash event $\crash$. 
Like durable linearizability, although crash events appear in the history, separate recovery operations do not explicitly appear in the histories. 
Invocation and
response events of the same operation are said to {\em match}.  Events
are furthermore parameterised by thread or transaction identifiers
from a set $\Tids$. For simplicity, we do not distinguish between
threads and transactions here.  Invocation events may have input
parameters and response events output parameters.  We use the
following notation on histories: for a history $h$, $h \zproject t$ is
the projection onto the events of transaction or thread $t$ only, and
$h[i..j]$ the subsequence of $h$ from $h(i)$ to $h(j)$ inclusive. We write $h h'$ for the concatenation of two histories $h$ and $h'$. We
say that two histories $h$ and $h'$ are {\em equivalent}, denoted $h
\equiv h'$, if $h\zproject t = h' \zproject t$ for all $t \in \Tids$.
For a response event $\ev$, we let $rval(\ev)$ denote the value returned
by $\ev$. If $\ev$ is not a response event, then we let $rval(\ev) = \bot$.
We furthermore let $Res$ be the set of all response and $Inv$ the set
of all invocation events.

We consider two types of histories, transactional and
non-transactional histories.  A transactional history only contains
the invocation and response events of
Table~\ref{tab:stmevents}. STM algorithms
allow for a concurrent access to shared memory (a set of locations
$\Loc$). Every transaction consists of an operation $\tmbegin$ followed
by a number of operations $\tmwrite$ or $\tmread$ and finally an
operation $\tmend$. All of these operations may also return ${\tt abort}$ meaning that the operation has not succeeded.  We say that a
transaction $t$ is {\em committed} in a history $h$ if
$res_t(\tmendc)$ is contained in $h$.  In non-transactional histories
we only have invocations and responses of operations on an object (e.g., a data structure), where invocations appear before their corresponding responses, and there is no grouping
of operations into transactions.

\begin{table}[t]
\begin{tabular}{l@{\qquad\qquad}l}
  invocations~ & possible matching responses \\
  \hline 
  $inv_t(\tmbegin)$ & $res_t(\tmbegino)$, $res_t(\tmbegina)$ \\[1pt]
  $inv_t(\tmend)$ & $res_t(\tmendc)$, $res_t(\tmenda)$ \\[1pt]
  $inv_t(\tmread(x))$ & $res_t(\tmread(v))$, $res_t(\tmreada)$ \\[1pt]
  $inv_t(\tmwrite(x,v))$ & $res_t(\tmwriteo)$, $res_t(\tmwritea)$ 
\end{tabular} 
\caption{Events appearing in transactional histories, where $t \in \Tids$ is
  a transaction identifier, $x \in \Loc$ is a location, and $v\in \Val$ a
  value}
\label{tab:stmevents}
\end{table} 

A (non-transactional) history is {\em sequential} if every invocation
event (except for possibly the last event) is directly followed by its
matching response.  A transactional history is {\em transaction
  sequential} if it is sequential and there are no overlapping
transactions.  A history is {\em complete} if there are no pending
operations, i.e., no invocations without a matching return.  The
function $complete$ removes all pending operations from a history.  A
history is {\em well-formed} if $h \zproject t$ is sequential for
every $t \in \Tids$.  A well-formed transactional history is
furthermore {\em transaction well-formed} if for every $t$,
$h\zproject t = \lseq \ev_0, \ldots, \ev_m\rseq$ is a sequential
history such that $\ev_0 = inv_t(\tmbegin)$, and for all $0 < i \leq
m$, event $\ev_i \neq inv_t(\tmbegin)$ and for all $0 < i < m$,
$rval(\ev_i) \notin \{\texttt{commit}, \texttt{abort}\}$. This, in
particular, implies that transaction identifiers cannot be re-used.

For a history $h$ and events $e_1, e_2$\footnote{Following Herlihy and
  Wing \cite{HeWi90}, we assume events are unique in a history by
  equipping them with a unique tag. For simplicity, these details are
  elided in our formalisation.}, we write (1) $e_1 <_h e_2$ whenever
$h= h_0 e_1 h_1 e_2 h_2$, and (2) $e_1 \ll_h e_2$ if $e_1 <_h e_2$ and
$e_1 \in Res, e_2 \in Inv$ (real-time order of operations).  In a
transactional history $h$, we furthermore write $t_1 \prec_h t_2$ if
the commit operation of transaction $t_1$ completes before transaction
$t_2$ starts (real-time order of transactions).

\subsection{Linearizability and Durable Linearizability}
\label{sec:lin-dlin}
Linearizability of concurrent data structures is defined by comparing
the (possibly concurrent) histories arising in usages of the data
structure to sequential {\em legal} histories.  Legality is defined by
specifying sequential objects $\bbS$, i.e., sequential versions of a
data structure (see \refdef{def:seq-obj}).  These sequential versions
define the ``correct'' behaviour, e.g. a queue data structure
adhering to a FIFO protocol or not losing elements.  For now, in the formal
definition of linearizability, we simply assume that we are given the 
set of sequential legal histories $H_\bbS$ (as generated by a  sequential 
object). Later, we will give an abstract data type in the form
of an IOA~\cite{LT87} as a specification of a sequential object.

A concurrent data structure is {\em linearizable} if all of its
histories arising in usages of the data structure are linearizable.

\begin{definition}[Linearizability~\cite{HeWi90}]
  A (concurrent) history $h$ is linearizable (w.r.t.~some set of sequential histories $H_\bbS$) iff
  there exists some $h_0 \in Res^*$ (completing some of the pending operations) such that 
  for $h'=complete(h h_0)$ there exists some $h_s \in H_\bbS$ such that
  \begin{description}
     \item[L1] $h' \equiv h_s$ and 
     \item[L2] $e \ll_{h'} e'$ implies $e \ll_{h_s} e'$. 
  \end{description} 
\end{definition}

\noindent In this definition, we assume the history does not contain any crash events. 
Linearizability only considers executions of data structures without intervening 
system crashes.

For durable linearizability, 
we need to consider histories with crash events.  Given a
history $h$, we let $\ops(h)$ denote $h$ restricted to non-crash
events.  The crash events partition a history into
$h = h_0 \crash_1 h_1 \crash_2 ...h_{n-1} \crash_n h_n$, such that $n$
is the number of crash events in $h$, $\crash_i$ is the $i$th crash event
and $ops(h_i) = h_i$ (i.e., $h_i$ contains no crash events).  We call the subhistory
$h_i$ the {\em $i$-th era} of $h$.  For well-formedness of histories
we now also require every thread identifier to appear in
at most one era.

These definitions allow us to lift linearizability to durable linearizability. 

\begin{definition}[Durable Linearizability~\cite{DBLP:conf/wdag/IzraelevitzMS16}] \label{def:d-lin}A history $h$ is {\em durably
    linearizable} iff it is well formed and $\ops(h)$ is linearizable.
\end{definition}

\noindent Durable linearizability will later be used to establish correctness of a library implementation that provides synchronised access to shared memory in the presence of  NVM. 

\subsection{Opacity and Durable Opacity} 

Opacity~\cite{2010Guerraoui,GuerraouiK08} compares concurrent
histories generated by an STM implementation to sequential histories. 
The difference to linearizability is that we  need to (a) consider entire transactions 
and (b) deal with aborted transactions. 
The correctness criterion opacity guarantees that values written by aborted transactions 
(i.e., transactions with events with $\abort$ as response value) cannot be read by other transactions. 

For opacity, we again compare concurrent histories against a set of
legal sequential ones, but now we employ transaction sequential
histories.  Again, we assume the set of legal sequential transactional
histories $TH_\bbS$ to be given, and out of these define the {\em
  valid} ones.

\begin{definition}[Valid History]
  \label{legal}
  Let $hs$ be a sequential history and $i$ an index of $hs$.  
  Let $hs'$ be the projection of $hs[0..(i-1)]$ onto all events of committed
  transactions plus the events of the transaction to which $hs(i)$
  belongs. 
  Then we say \emph{$hs$ is valid at $i$} whenever $hs'$ is legal. 
  We say \emph{$hs$ is valid} iff it is valid at each index
  $i$. 
\end{definition}

\noindent We let $VH_\bbS$ be the set of transaction sequential valid histories. 
With this at hand, we can define opacity similar to linearizability. 

\begin{definition}[Opacity~\cite{GuerraouiK08,2010Guerraoui}]
  \label{def:opaque}  
  A (concurrent) history $h$ is {\em end-to-end opaque} iff there exists some
  $h_0 \in Res^*$ (completing some of the pending operations) such that 
  for $h'=complete(h h_0)$ there exists some $h_s \in VH_\bbS$ such that
  \begin{description}
    \item[O1] $h' \equiv h_s$, and 
    \item[O2] $t_1 \prec_{h} t_2$ implies $t_1 \prec_{h_s} t_2$. 
  \end{description} 
  A history $h$ is {\em opaque}
  iff each prefix $h'$ of $h$ is end-to-end opaque. 
\end{definition}

\noindent An STM algorithm itself is \emph{opaque} iff its set of
histories occuring during executions of the STM is opaque. For durable
opacity, we simply lift this definition to histories with crashes.

Like durable linearizability, the purpose of durable opacity is to
ensure that histories with crashes leave the shared state in a
consistent state as defined by opacity. This means that any live
transaction that has not yet started its commit operation will be
treated as an aborting transaction. If a transaction has started its
commit, then the commit could be completed by either a successful
commit or an abort. 
For well-formedness of TM histories,
we now also require every {\em transaction} identifier to appear in
at most one era. This means that no
transaction survives a crash.
  
\begin{definition}[Durable Opacity~\cite{DBLP:conf/forte/BilaDDDSW20}]
  \label{def:duropaque}
  A history $h$ is {\em durably opaque} iff it is transaction well-formed
  and $ops(h)$ is opaque.
\end{definition}

One of the guarantees of durable opacity (again like durable
linearizability) is that it ensures every committed transaction is
persisted. Thus, if a transaction's effects are globally visible, then
this transaction is guaranteed to also survive any subsequent
crashes. In other words, durable opacity ensures that for committed transactions, the {\em visibility
order} (the order in which transactions are seen by other transactions) and the {\em persistent order} (the order in which transactions become durable) coincide.

Furthermore, durable opacity aims to transfer the atomicity property of opacity to the NVM setting. For opacity, this property has been shown via a study of a specification called TMS1~\cite{DGLM13}. It is well known that TMS1  is both necessary and sufficient to ensure transactions are atomic~\cite{DBLP:conf/wdag/AttiyaGHR14}. Opacity is known to be stronger than TMS1~\cite{LLM12}, thus also guarantees the sufficiency property. Durable opacity ensures transactional atomicity even in the presence of crashes, thus ensures the same guarantees. However, the precise formulation of the atomicity problem in the setting of NVM deserves further study.  

In this paper, we aim to develop a method for proving durable opacity
of STM algorithms. For the proof, we develop a modular proof technique,
which requires us to show durable linearizability of some library data
structure providing access to shared memory.

\section{Using IOA to Prove Durable Opacity} 
\label{sec:prov-opac-meth}

Previous works~\cite{DBLP:conf/forte/ArmstrongDD17,DBLP:conf/opodis/DohertyDDSW16,DBLP:journals/csur/DongolD15,DBLP:conf/forte/ArmstrongD17}
have considered proofs of opacity using the operational \TMS
specification~\cite{DGLM13}, which has been shown to guarantee opacity~\cite{LLM12}. The proofs show refinement of the implementation against
the \TMS specification using either forward or backward simulation. In
this, both implementation and specification are given as Input/Output automata (IOA) to enable use of a standard simulation-based proof technique.  For durable opacity, we
follow a similar strategy.  We develop the \DTMS operational
specification, a durable version of the \TMS specification, that we
prove satisfies durable opacity.  By proving a simulation relation to
hold between a (durable) STM implementation and \DTMS we can establish
durable opacity of an STM.

In the following, we will first of all shortly explain IOA and simulations in general 
and thereafter develop \DTMS.

\subsection{IOA, Refinement and Simulation}
We use Input/Output Automata (IOA) \cite{LT87} to model both STM 
implementations and the specification, \DTMS.
\begin{definition}[Input/Output Automaton (IOA)]
  An \emph{Input/Output Automaton (IOA)} is a labeled transition
  system $A$ with a set of \emph{states} $states(A)$, a set of
  \emph{actions} $acts(A)$, a set of \emph{start states}
  $start(A)\subseteq states(A)$, and a \emph{transition relation}
  $trans(A)\subseteq states(A)\times acts(A)\times states(A)$ (so that
  the actions label the transitions).
\end{definition}
The set $acts(A)$ is partitioned into input actions $input(A)$, output
actions $output(A)$ and internal actions $internal(A)$.  The internal
actions represent events of the system that are not visible to the
external environment. The input and output actions are externally
visible, representing the IOA's interactions with its
environment. Thus, we define the set of {\em external actions},
$external(A) = input(A)\cup output(A)$. We write
$s \stackrel{a}{\longrightarrow}_A s'$ iff $(s, a, s') \in trans(A)$.

An {\em execution} of an IOA $A$ is a
sequence $\sigma = s_0 a_0 s_1 a_1 s_2 \dots s_n a_n s_{n+1}$ of
alternating states and actions,  such that $s_0 \in start(A)$ and 
for all states $s_i$, $s_i \stackrel{a_{i}}{\longrightarrow}_A s_{i+1}$.
We write $\exec(A)$ for the set of all executions of $A$ and
$\firststate(\sigma) = s_0$ for the initial state of an
execution $\sigma$. Whenever we have several IOAs, we use
indices to distinguish between them, e.g.~$\sigma_A$ is
used to denote an execution of $A$. 

A {\em reachable} state of $A$ is a state appearing in an execution of $A$. 
We let $reach(A)$ denote the set of all reachable states of $A$.
An {\em  invariant} of $A$ is any superset of the reachable states of $A$
(equivalently, any predicate satisfied by all reachable states of
$A$). A {\em trace} of $A$ is any sequence of (external) actions
obtained by projecting the external actions of any execution of
$A$. The set of traces of $A$, denoted $traces(A)$, represents $A$'s
externally visible behaviour.

For IOA $C$ and $A$, we say that $C$ is a
{\em refinement} of $A$, denoted $C \leq A$,  iff $traces(C) \subseteq traces(A)$.
Note that refinement is transitive. We typically show that $C$ is a refinement
of $A$ by proving the existence of a {\em forward simulation}, which
enables one to check step correspondence between the transitions of
$C$ and those of $A$. The definition of forward simulation we use is
adapted from that of Lynch and Vaandrager~\cite{LynchVaan95}.

\noindent
\begin{minipage}[t]{\columnwidth}
\begin{definition}[Forward Simulation]
\label{def:for-sim}
A \emph{forward simulation} from a concrete IOA $C$ to an abstract IOA
$A$ is a relation $R \subseteq states(C) \times states(A)$ such that 
each of the following holds.  \smallskip

\noindent \emph{Initialisation}. $\forall cs \in start(C).\ \exists as \in start(A).\ R(cs, as)$\smallskip

\noindent \emph{External step correspondence}.

\hfill $\begin{array}[t]{@{}l@{}}
  \forall cs \in reach(C), as \in reach(A), a \in external(C), cs' \in states(C).\ \\
  \qquad R(cs, as) \wedge cs \stackrel{a}{\longrightarrow}_C cs' \imp \exists as' \in states(A).\ R(cs', as') \wedge as \stackrel{a}{\longrightarrow}_A as'
\end{array}
$ \hfill {}\smallskip

\noindent \emph{Internal step correspondence}.

\hfill
$\begin{array}[t]{@{}l@{}}
   \forall cs \in reach(C), as \in reach(A), a \in internal(C), cs' \in states(C).\ \\
   \qquad R(cs, as) \wedge cs \stackrel{a}{\longrightarrow}_C cs' \imp \\
   \qquad \begin{array}[t]{@{}l@{}}
     R(cs', as) \lor  
     \exists a' \in internal(A), as' \in states(A).\ R(cs', as') \wedge as \stackrel{a'}{\longrightarrow}_A as'
   \end{array}
 \end{array}
 $\hfill{}
\end{definition}
\end{minipage}
Forward simulation is {\em sound} in the sense that if there is a
forward simulation between $A$ and $C$, then $C$ refines $A$ \cite{LynchVaan95,MullerIOA1998}.

\subsection{Canonical IOA for (durable) linearizability}
\label{sec:canonical}

To prove linearizability the relevant set of sequential
histories $H_\bbS$ are given as the histories of a sequential
object $\bbS$, that defines a set atomic operations $op_i$ that
receive input, modify a state and return output.  

\begin{definition}[Sequential Object]\label{def:seq-obj}
  A \emph{sequential object} $\bbS$ is a 4-tuple $(\Sigma,Val, State, Init)$ where
  \begin{itemize}
  \item $State$ is a set of states, $Init \subseteq State$ is a set of initial states,
  \item $Val$ is a set of values used as input and output,
    \item $\Sigma$ is a set of atomic operations $op_i$ for some $i \in I$.\\
      Each operation is specified as a relation
      $op_i \subseteq Val \times State \times State \times Val$.
  \end{itemize} 
\end{definition}

Some operations may have no inputs/outputs and others may have several. This can
be accommodated by including tuples including the empty tuple $\epsilon$
in $Val$. We drop an empty input or output when writing an event.
A sequential history of $\bbS$ has the form
\begin{align*} 
inv(op_{k_1}(in_1)), res(op_{k_1}(out_1)), \ldots, inv(op_{k_n}(in_n)), res(op_{k_n}(out_n))
\end{align*}

\noindent The history is a legal sequential history in $H_S$,
iff there is a sequence $s_0 \ldots s_n$ of
states, such that $s_0 \in Init$ and $(in_m, s_m, s_{m+1}, out_m) \in op_{k_m}$
for all $m < n$. 

To prove durable linearizability of a concurrent implementation we
will specify the concurrent program as an IOA $C$ that generates a set
of concurrent histories.  Note that--as to mimic execution of an NVM
architecture--this implementation IOA $C$ would need to explicitly
model persistent and volatile memory as well as its flushing
discipline, i.e., when the implementation wants an update to a
location to reach persistent memory.
To prove that $C$ is durably linearizable to $H_\bbS$, 
it is then sufficient to prove that $C$ refines
the {\em canonical durable IOA} $\duraut{\bbS}$
shown in Fig.~\ref{fig:canonical} (see~\cite{DBLP:conf/fm/DerrickDDSW19}). 
This IOA serves as an abstract specification of durable
linearizability in the refinement proof: its traces are exactly
the durably linearizable histories (of some sequential object). 

The state of this IOA incorporates the state $s$ of the
sequential object $\bbS$ and adds
a program counter $pc_t$ for every transaction $t \in T$.
The possible values of this program counter include {\em notStarted},
{\em ready} and {\em crashed} to indicate that the transaction has not
started (its initial value), is running but not currently
executing an operation, or has crashed. 
The execution of an operation $op$ is split into three steps: an invocation and a response 
of the operation plus a $do$-step (where the actual effect of the operation takes place). 
Note that both $run$ and $do$ are internal actions and thus do not appear in the traces of the IOA. 
\begin{itemize}
\item First, when  $pc_t$ = {\em ready}
     an invoke step with action $inv_t(op(in))$ is executed.
     The input value of this step is arbitrary and gets
     stored in  $pc_t$ by setting it to $doOp(in)$.
\item Second, a step with internal action $do_t(op)$
     is executed. This step will
      correspond to the linearization point of an implementation.
      The step modifies the state 
      of $op$ by choosing a new state and an output
      according to the specification of $op$ (the step is not
      possible if there is no $s',out'$ with $op(in, s,s', out')$).      
      The computed output is again stored in
      $pc_t$ by setting it to $resOp(out)$. 
\item Finally, a response step, that returns the $out$
      value that was stored in $pc_t$ by emitting
      an action $resOp(out)$. This step finishes
      the execution of $op$ by setting $pc_t$ to {\em ready}.
\end{itemize}

The durable canonical IOA (more details are in \cite{DBLP:conf/fm/DerrickDDSW19})
is an extension of the canonical IOA from
\cite{DBLP:books/mk/Lynch96} for linearizability
to accommodate durable linearizability. It is the most general
specification of concurrent runs that
still allows us to construct an equivalent sequential history:
the sequential history can be constructed as a sequence
of invoke-response pairs from the sequence
of executed $do_t(op)$ steps. The IOA guarantees
that this sequential history is obviously in $H_\bbS$.

The following theorem establishes a correspondence between the durable
IOA and durable linearizability. For a sequential object $\bbS$,
we let $\durlin{\bbS}$ be the set of histories that are durably
linearizable with respect to $\bbS$.

\begin{thmC}[\cite{Derrick+21}]
Let $\bbS$ be a sequential object. Then $traces(\duraut{\bbS}) = \durlin{\bbS}$.
\end{thmC}

As the durable IOA has durably linearizable histories only, it can serve as an abstract specification in a proof of durable linearizability via refinement. 

\begin{lemma}\label{durlin-equiv-refine-canonical}
  Let $C$ be an implementation IOA. 
  If $C$ refines $\duraut{\bbS}$, then $C$ is durably linearizable to $\bbS$.  
\end{lemma}

Summarising, this gives us the following: Whenever we have an
algorithm $Alg$ which runs on an NVM architecture and the
implementation IOA $C$ models the executions of this algorithm
on NVM (i.e., adequately represents persistent and volatile memory)
and $C$ refines $\duraut{\bbS}$, then the algorithm $Alg$ is durably
linearizable.

\begin{figure}[!t]
  \small 
\newlength{\myrowsep}
\setlength{\myrowsep}{5em}  
\begin{tabular}{lll} 
  \action{inv_t(op(in))}
  {pc_t = \pcReady}
  { pc_t := doOp(in)}  & 
  \action{do_t(op)}
  {pc_t = doOp(in) }
  { (s,out) := SOME(s', out').\\
    & \qquad\quad\qquad op(in,s,s', out') \\
     & pc_t := resOp(out)} & 
  \action{\resp_t(op(out))}
  {pc_t = resOp(out)}
  { pc_t := \pcReady} \\[\myrowsep]
  \action{run_t}
  {pc_t = \pcNotStarted }
  { pc_t := \pcReady }  & 
  \action{crash}
  { true }
{pc := \lambda t: T. \\
       & \quad \textbf{if\ } pc_t \neq \pcNotStarted \\
       & \quad \textbf{then\ } \pcCrashed    \textbf{\ else\ } pc_t 
  } & 
\end{tabular}

\caption{Durable IOA $\duraut{\bbS}$}
\label{fig:canonical}
\end{figure}

\subsection{Refinement in context}
\label{sec:ref-in-context}
For our modular proof technique, we will let an STM algorithm  
call a library in order to manage access to shared state. 
As both STM and library will be formalised in terms of an IOA, 
we need some notion of a {\em context} IOA (i.e., the STM IOA) using a {\em library IOA}. 
To this end, we employ the following definition of product IOA~\cite{DBLP:books/mk/Lynch96}, 
which requires synchronisation of two IOA on shared external actions. 

\begin{definition}[Product IOA]\label{product}
  Let A, B be two IOA with no shared internal actions.
  Then the {\em product  IOA} $A \times B$ is defined to have
  \begin{itemize}
   \item states($A \times B$) =  $states(A) \times states(B)$,
   \item start($A \times B$) = $start(A) \times start(B)$,
   \item acts($A \times B$) = $acts(A) \union acts(B)$, 
   \item $(as,bs) \stackrel{a}{\longrightarrow}_{A\times B} (as',bs')$ iff the following two properties hold:\\
     if $a \in actions(A)$, then $as \stackrel{a}{\longrightarrow}_A as'$, else $as' = as$; \\
     if $a \in actions(B)$, then $bs \stackrel{a}{\longrightarrow}_B bs'$, else $bs' = bs$.
  \end{itemize}
\end{definition}

\noindent In the following we will use this product construction in an
asymmetric way: the shared external actions of $A$ (the library)
and $\LB{\cdot}$ (the STM algorithm) are the invocations and responses of
library calls, and the library is required to have no
further external actions.  In such a setting, we write $\LB{A}$ for the
product of $A$ and $\LB{\cdot}$ (i.e., where IOA $\LB{\cdot}$ uses library IOA $A$).

Later we will develop two versions of the library which provides
access to shared memory: one with and one without volatile
memory. These two versions are shown to be a refinement of each other
(more precisely, one version is shown to be durably linearizable
w.r.t.~the other), and we need to lift this result to STMs using the
libraries.  It is folklore knowledge that refinement of an abstract
object by a concrete object implies refinement between an algorithm (a
context) using the abstract object and the same algorithm using the
concrete one.  Theorems stating such a property have been proven in
many settings, e.g.~for data refinement in~\cite{de1998data}.  The
fundamental paper on linearizability \cite{HeWi90} implicitly uses
such a result when it assumes that the individual steps of algorithms
are linearizable operations as well.  We could however not find a
formal proof of refinement in context for IOA, and thus both state and prove it in this setting.

\begin{theorem} \label{th:ref-context} If $C \le A$, then $\LB{C} \le \LB{A}$.
\end{theorem}
\noindent Note that $C \le A$ and $\LB{C} \le \LB{A}$ implies that the external actions of $C$ and $A$ are the same and that they are a subset of external actions of $B$.

To prove refinement, we need to construct an execution $\sigma_{\LB{A}}$
of $\LB{A}$ with the same trace (i.e., the same external events) when
given an execution $\sigma_{\LB{C}}$ of $\LB{C}$.  The idea is shown in
Fig.~\ref{product-refine} with an execution that executes three events
$a_1, a_2, a_3$.  The execution of $\LB{C}$ contains an execution
$\sigma_C$ of $C$ by projecting to the states of $C$ and removing all
steps (here: $a_1$) where $C$ is not involved.  This execution can be
split into finite segments of internal $C$-steps that each end with an
external shared action (with possibly a final sequence of internal
$C$-steps that is not used).  In the example, there is one segment
consisting of one internal action $a_2$, ending with the external
action $a_3$.  By refinement, there exists an execution $\sigma_A$ of A
with the same external actions. This execution can be split in the
same way: in the example the new segment consists of internal actions $\alpha$, and
ends with $a_3$.  Now delete the internal $C$-steps from the combined
execution of $\LB{C}$, and replace the $C$-step in each combined step of
$C$ and $B$ with the corresponding $A$-step from the abstract
execution.  Add the sequence of internal $A$-steps (here: $\alpha$)
that leads to this step (here: $a_3$) right before the combined step
in the combined execution.  The result is an execution $\sigma_{\LB{A}}$
of $\LB{A}$ which has the same trace as the original execution.  Formally,
the two steps are done by a projection function
$\pi_C$ and a $\replfun$ function.

\begin{figure}
  \centering
  \scriptsize
  \begin{tikzpicture}[node distance=2cm,auto]
    \node (ba0) [inner sep=2pt]{$(bs_0, as_0)$};
    \node (a0) [below=1cm of ba0, inner sep=2pt]{$as_0$};
    \node (c0) [below=1cm of a0, inner sep=2pt]{$cs_0$};
    \node (bc0) [below=1cm of c0, inner sep=2pt]{$(bs_0, cs_0)$};
    
    \node (ba1) [right=1.5cm of ba0, inner sep=2pt]{$(bs_1, as_0)$};
    \node (bc1) [right=1.5cm of bc0, inner sep=2pt]{$(bs_1, cs_0)$};
    
    \node (ba2) [right=1.5cm of ba1, inner sep=2pt]{$(bs_1, as_1)$};
    \node (a2) [below=1cm of ba2, inner sep=2pt]{$as_1$};
    \node (c2) [below=1cm of a2, inner sep=2pt]{$cs_1$};
    \node (bc2) [right=1.5cm of bc1, inner sep=2pt]{$(bs_1, cs_1)$};
    
    \node (ba3) [right=1.5cm of ba2, inner sep=2pt]{$(bs_2, as_2)$};
    \node (a3) [below=1cm of ba3, inner sep=2pt]{$as_2$};
    \node (c3) [below=1cm of a3, inner sep=2pt]{$cs_2$};
    \node (bc3) [right=1.5cm of bc2, inner sep=2pt]{$(bs_2, cs_2)$};
    
    \node (bax) [right=0.5cm of ba3, inner sep=3pt]{...};
    \node (ax) [right=0.85cm of a3, inner sep=3pt]{...};
    \node (cx) [right=0.85cm of c3, inner sep=3pt]{...};
    \node (bcx) [right=0.5cm of bc3, inner sep=3pt]{...};
    
    \node (execba) [right=0cm of bax, inner sep=2pt]{$\in \exec(\LB{A})$};
    \node (execa) [right=0cm of ax, inner sep=2pt]{$\in \exec(A)$};
    \node (execc) [right=0cm of cx, inner sep=2pt]{$\in \exec(C)$};
    \node (execbc) [right=0cm of bcx, inner sep=2pt]{$\in \exec(\LB{C})$};

    \path[->] (ba0) edge node{$a_1$} (ba1);
    \path[->] (ba1) edge node{$\alpha$} (ba2);
    \path[->] (ba2) edge node{$a_3$} (ba3);
    \path[->] (ba3) edge (bax);
    
    \path[->] (a0) edge node{$\alpha$} (a2);
    \path[->] (a2) edge node{$a_3$} (a3);
    \path[->] (a3) edge (ax);
       
    \path[->] (c0) edge node{$a_2$} (c2);
    \path[->] (c2) edge node{$a_3$} (c3);
    \path[->] (c3) edge (cx);
    
    \path[->] (bc0) edge node{$a_1$} (bc1);
    \path[->] (bc1) edge node{$a_2$} (bc2);
    \path[->] (bc2) edge node{$a_3$} (bc3);
    \path[->] (bc3) edge (bcx);
    
    \path[->] (execbc) edge[bend left=20] node{$\pi_C$} (execc);
    \path[->] (execc) edge[bend left=20] node{refine} (execa);
    \path[->] (execbc) edge[bend right=30, transform canvas={xshift=3mm}] (execba);
    \path[->] (execa) edge[bend left=30, transform canvas={xshift=2mm}] node[right]{$\replfun$} (execba);
    
  \end{tikzpicture}
  \caption{Construction of an execution of \LB{A} from an execution of \LB{C}. $a_1 \in acts(B) \setminus external(C)$, $a_2 \in  interal(C)$, $a_3 \in external(C)$ ($= external(A)$, $\subseteq external(B)$), $\alpha \in internal(A)^*$. }\label{product-refine}
\end{figure}

\smallskip 
\begin{proof}[Proof of~\autoref{th:ref-context}:]
  To show that $traces(\LB{C}) \subseteq traces(\LB{A})$ choose an arbitrary trace from
  $traces(\LB{C})$. For this trace an execution  
  $\sigma_{\LB{C}}$ of the form $(bs_0,cs_0)\;a_0\;(bs_1,cs_1) \ldots$ $a_n\;(bs_{n+1},cs_{n+1})$ must exist.  
  For such an execution the projection $\pi_C: \exec(\LB{C}) \rightarrow \exec(C)$
  to an execution of C can be defined recursively over its length $n$.
  \[
  \pi_C((bs_0,cs_0)) = cs_0\\
  \pi_C((bs_0,cs_0)\;a_0\;\sigma'_{\LB{C}}) = cs_0\;a_0\;\pi_C(\sigma'_{\LB{C}})\ \text{when}\ a_0 \in acts(C)\\
  \pi_C((bs_0,cs_0)\;a_0\;\sigma'_{\LB{C}}) = \pi_C(\sigma'_{\LB{C}}) \ \text{when}\ a_0 \in acts(B) \setminus external(C)
  \]

  In the second and third line $\sigma'_{\LB{C}}$ is the rest of trace (of length $n$) with the first state and action removed.
  The actions of $\pi_C(\sigma_{\LB{C}})$ are those of $\sigma_{\LB{C}}$ which are in acts(C).
  Refinement then guarantees the existence of an execution $\sigma_A \in exec(A)$ with
  $\trace(\pi_C(\sigma_{\LB{C}})) = \trace(\sigma_A)$.

This allows to define a function $\replfun: \exec(\LB{C}) \times \exec(A) \rightarrow \exec(\LB{A})$.
The result of $\repl{\sigma_{\LB{C}}}{\sigma_A}$ is defined when $trace(\pi_C(\sigma_{\LB{C}})) = trace(\sigma_A)$.
It replaces the steps of C in $\sigma_{\LB{C}}$ with the corresponding ones in $\sigma_A$. Again, $\replfun$
is defined recursively.
For $n = 0$ we simply have
\begin{align}\label{repl0}
&\repl{(bs_0,cs_0)}{\sigma_A} = (bs_0, \firststate(\sigma_A))&
\end{align}
\noindent
When $n > 0$ there are two cases: When $a_0 \in Acts(B) \setminus external(C)$ then
\begin{align}\label{repl1}
& \repl{(bs_0,cs_0)\;a_0\;\sigma'_{\LB{C}}}{\sigma_A} = (bs_0, \firststate(\sigma_A))\;a_0\;\repl{\sigma'_{\LB{C}}}{\sigma_A} &
\end{align}
\noindent
Note that the first component of the pair $\firststate(\sigma'_{\LB{C}})$ is $bs_0$ in this case,
and that $trace(\pi_C(\sigma'_{\LB{C}})) = trace(\sigma_A)$ is still true for the recursive call.
Otherwise, when $a_0 \in external(C)$ then $a_0$ is in $external(A)$ and $external(B)$ as well, since the external
actions of C and A are the same and shared with B.
The trace of $\pi_C(\sigma_{\LB{C}})$ then contains $a_0$ as its first external action. 
By trace equality, the first external action of $\sigma_A$ must be $a_0$ as well. $\sigma_A$
therefore has the form $\sigma_A = as_0\;a'_1\;as_1\;a'_2\;\ldots\;a'_m\;as_m\;a_0\;\sigma'_A$
where $m \ge 0$, and $a'_1\;\ldots\;a'_m \in internal(A)^*$
is a sequence of internal actions. The sequence $\alpha$ in the example
of Fig.~\ref{product-refine} is this sequence of actions.
The resulting execution now first executes the internal steps, and finally the combined step, so
$\replfun$ is defined in this case as
\begin{align}\label{repl2}
&\repl{(bs_0,cs_0)\;a_0\;\sigma'_{\LB{C}}}{\sigma_A} =& \notag\\
&\qquad (bs_0,as_0)\;a'_1\;(bs_0,as_1)\;\ldots\;a'_m\;(bs_0,as_m)\;a_0\;\repl{\sigma'_{\LB{C}}}{\sigma'_A} &
\end{align}
\noindent
Again, $trace(\pi_C(\sigma'_{\LB{C}})) = trace(\sigma'_A)$ is still true for the recursive call.
It is now easy to check inductively that $\repl{\sigma_{\LB{C}}}{\sigma_A}$ returns an execution
  of $\LB{A}$, since all steps of the constructed execution are steps of $\LB{A}$.The first step of (\ref{repl1}) is a step of B with an unshared action of B that does not change the state of A, so it is a step of the product B[A] (last clause of Definition \ref{product}). The first m steps of (\ref{repl2}) are steps with unshared internal actions of A that do not change the state of B, so they are steps of B[A] too. Finally, step m+1 of (\ref{repl2}) executes shared action $a_0$ and changes both states according to the definition of A and B, so it is a step of B[A] too.
  
  When $\firststate(\sigma_{\LB{C}}) = (bs_0, cs_0)$, then the first state of
  $\repl{\sigma_{\LB{C}}}{\sigma_A}$ is $(bs_0,\firststate(\sigma_A))$,
which is initial, if the first states of $\sigma_{\LB{C}}$ and $\sigma_A$ are. 
The result $\repl{\sigma_{\LB{C}}}{\sigma_A}$  has the same trace as $\sigma_{\LB{C}}$ since all external
actions are preserved. This implies that the original $trace(\sigma_{\LB{C}})$ the construction started with
is also a trace of $\LB{A}$, finishing the proof.
\end{proof}

\smallskip

\noindent
{\bf Remark:} Although we do not need this generalisation here,
the result holds as well if refinement is defined as trace inclusion for finite \emph{as well as infinite} traces
(see~\cite{LynchVaan95}). Both $\pi_C$ and $\replfun$ are prefix-monotone,
so the result of applying the functions to infinite traces can be defined as the limit
of applying them to finite prefixes.
The proof for this extended scenario has been formalised
in KIV \cite{KIV-DNOREC}.

\subsection{IOA for \DTMS}
\label{sect:TMS2-def}

In this section, we describe the \DTMS specification, an operational
model that ensures durable opacity, which is based on
\TMS~\cite{DGLM13}. \TMS itself has been shown to imply
opacity \cite{LLM12}, and hence has been widely used as an
intermediate abstract specification in the verification of transactional memory
implementations~\cite{DBLP:conf/forte/ArmstrongDD17,DDSTW15,DBLP:conf/forte/ArmstrongD17,DBLP:conf/opodis/DohertyDDSW16}. \DTMS is thus designed to play the r\^{o}le of an abstract specification for refinement proofs of durable opacity like $\duraut{\bbS}$ is for proofs of durable linearizability. 

In the following, we let $f \oplus g$ denote functional override of $f$ by $g$, where we define
$f \oplus g = \lambda k \in \dom(f).\ {\bf if}\ k \in \dom(g)\
{\bf then}\ g(k) \ {\bf else}\ f(k)$.

\begin{figure}
\raggedright
{\small
{\bf State variables:}\\
$mems : seq (\Loc \to \Val)$, initially satisfying $\dom mems = \{0\}$\\  
$pc_t : PCVal$, for each $t\in T$, initially $pc_t=\pcNotStarted$ for all $t\in T$\\
$beginIdx_t : \nat$ for each $t\in T$, unconstrained initially\\
$rdSet_t: \Loc \pfun \Val$, initially empty for all $t\in T$, where $\pfun$ denotes a partial function\\
$wrSet_t: \Loc \pfun \Val$, initially empty for all $t\in T$\\
\centering
{\bf Transition relation:} \\[1em]
\bgroup
\setlength{\tabcolsep}{1em}
\setlength{\myrowsep}{4em} 
\newlength{\myrowsepa}
\setlength{\myrowsepa}{3em} 
\newlength{\myrowsepb}
\setlength{\myrowsepb}{5.5em} 
\begin{tabular}{@{}l@{\qquad\quad}l@{}}
\action{\beginInv{t}}
{pc_t = \pcNotStarted}
{pc_t := \pcBeginPending\\&
 beginIdx_t := len(mems) - 1}
&
\action{\beginResp{t}}
{pc_t = \pcBeginPending}
{pc_t := \pcReady}
\\[\myrowsep]
\action{\readInv{t}{l}}
{pc_t = \pcReady}
{pc_t := \pcDoRead(l)}
&
\action{\readResp{t}{v}}
{pc_t = \pcReadResp(v)}
{pc_t := \pcReady}
\\[\myrowsepa]
\action{\writeInv{t}{l, v}}
{pc_t = \pcReady}
{pc_t := \pcDoWrite(l, v)}
&
\action{\writeResp{t}}
{pc_t = \pcWriteResp}
{pc_t := \pcReady}
\\[\myrowsepa]
\action{\commitInv{t}}
{pc_t = \pcReady}
{pc_t := \pcDoCommit}
&
\action{\commitResp{t}}
{pc_t = \pcCommitResp}
{pc_t := \pcCommitted}
\\[\myrowsepa] 
\action{\abortResp{t}}
{pc_t \notin \{\pcNotStarted, \pcReady, \\
  & \pcCommitResp, \pcCommitted, \pcAborted\}}
{pc_t := \pcAborted}
  &
    \action{\doWrite{t}{l, v}}
    {pc_t = \pcDoWrite(l, v)}
    {pc_t := \pcWriteResp\\&
  wrSet_t := wrSet_t\oplus \{l \to v\}
  }
  \\[\myrowsep]
\action{\doCommitReadOnly{t}{n}}
{pc_t = \pcDoCommit\\&
 \dom(wrSet_t) = \emptyset\\&
 validIdx(t, n)}
{pc_t := \pcCommitResp}
&
\action{\doCommitWriter{t}}
{pc_t = \pcDoCommit\\&
 rdSet_t \subseteq last(mems)}
{pc_t := \pcCommitResp\\&
 mems := mems \cat (last(mems) \oplus wrSet_t)} 
\\[\myrowsepb]
\action{\doRead{t}{l, n}}
{pc_t = \pcDoRead(l)\\&
 l\in \dom(wrSet_t) \vee validIdx(t, n)}
{{\bf if}\ l \in \dom(wrSet_t)\\ 
&  {\bf then}\ pc_t := \pcReadResp(wrSet_t(l))\\&
 {\bf else}\ v := mems(n)(l)\\&
 \ \ \ \ \ \ \ \  pc_t := \pcReadResp(v)\\&
 \ \ \ \ \ \ \ \ rdSet_t := rdSet_t\oplus \{l \to v\}}
&
\begin{tabular}[t]{@{}l@{}}
  \action{crashRecovery}
  {true}
  {pc := \lambda t: \Tids. \\
  & \qquad {\bf if}\ pc_t \notin \{\pcNotStarted, \pcCommitted\} \\
  & \qquad {\bf then}\ \pcAborted \\
  & \qquad {\bf else}\ pc_t \\
  & mems := \langle last(mems) \rangle}
\end{tabular}
\end{tabular}
\egroup
\smallskip

{\bf where}
$\begin{array}[t]{r@{~~}c@{~~}l}
  PCExternal &\sdef& \{\pcNotStarted, \pcReady, \pcCommitResp,
                      \pcWriteResp, \pcCommitted, \pcAborted\}\cup {} \\
   & & \{\pcReadResp(v) \mid v\in \Val\}%
                      \\
   PCVal &\sdef & PCExternal \cup \{\pcBeginPending, \pcDoCommit, \pcCancelPending\}\\
         &&{} \cup \{\pcDoRead(l) \mid l\in L\} \cup\{\pcDoWrite(l, v) \mid l\in \Loc, v\in \Val\}\\[1mm]
  validIdx(t, n) &\sdef & beginIdx_t \leq n < len(mems) \wedge
                          rdSet_t \subseteq mems(n) \\
  op & \in & \{{\tt TMBegin}, {\tt TMRd}, {\tt TMWr}, {\tt TMCommit}\} 
\end{array}
$
}
\caption{The state space and transition relation of \DTMS, which
  extends \TMS with a crash-recovery event} 
  \label{fig:tms2}
\end{figure}

Formally, \DTMS is specified by the IOA in \reffig{fig:tms2}, which
describes the required ordering constraints, memory semantics and
prefix properties. 
Recall that we assume a set $\Loc$ of locations and a set $\Val$ of  
values.  
Thus, a memory is modelled by a function of type $\Loc \to
\Val$.   
A key feature of \DTMS (like TMS2) is that it keeps track of a
{\em sequence} of memory states, one for each committed writing
transaction. This makes it simpler to determine whether reads are
consistent with previously committed write operations. Each committing
transaction containing at least one write adds a new memory version to
the end of the memory sequence. 
Note that \DTMS is an IOA used for abstract specification in a refinement proof only;
it is not an implementation that has to keep track of persistent and volatile memory. 
 

The state space of \DTMS has several components. The first, $mems$ is
a nonempty sequence of {\em memory} states, which initially contains one state. 
The original specification of TMS2 is parameterised by some initialisation 
predicate describing this initial memory state, which we elide here for simplicity (and simply assume 
the implementation to employ the same initialisation).
For each transaction $t$ there is
a program counter variable $pc_t$, which ranges over a set of {\em
  program counter values}, which are used to ensure that each  
transaction is well-formed, and to ensure that each transactional
operation takes effect between its invocation and response. There is
also  a {\em begin index} variable $beginIdx_t$, that is set to the
index of the most recent memory version when the transaction
begins. This variable is critical to ensuring the real-time ordering
property between transactions.  Finally, there is a {\em read set},
$rdSet_t$, and a {\em write set}, $wrSet_t$, which record the values
that the transaction has read and written during its execution,
respectively. 

The read set is used to determine whether the values that have been
read by the transaction are consistent with the same version of memory
(using $validIdx$). The write set, on the other hand, is required
because writes in \DTMS are modelled using {\em deferred update}
semantics: writes are recorded in the transaction's write set, but are
not published to any shared state until the transaction commits.


The $crashRecovery$ action again models the effect of crashes and consecutive recoveries. 
It sets the program
counter of every in-flight transaction to $aborted$, which prevents
these transactions from performing any further actions in the era
following the crash (for the generated history). Note that since
transaction identifiers are not reused, the program counters of
completed transactions need not be set to any special value (e.g.,
$crashed$) as with durable
linearizability~\cite{DBLP:conf/fm/DerrickDDSW19}.  Moreover, after
restarting, it must not be possible for any new transaction to
interact with memory states prior to the crash. We therefore reset the
memory sequence to be a singleton sequence containing the last memory
state prior to the crash.

The external actions of \DTMS are all invocation and response actions ($inv_t$ and $res_t$) 
plus the new $crashRecovery$ action. The latter is the crash action $c$ 
of histories. Note that the traces of \DTMS hence take the form of histories. 

The following theorem ensures that \DTMS can be used as an
intermediate specification in our proof method.

\begin{theorem}
\label{theor:dtsm-sound}
Each trace of\, \DTMS is durably opaque.
\end{theorem}

 The proof of this theorem can be found in the appendix of~\cite{DBLP:conf/forte/BilaDDDSW20}. 
Like durable linearizability, we have the following lemma, which allows us to establish durable opacity using refinement.

\begin{lemma}
 Let $C$ be an implementation IOA. If $C$ refines \DTMS, then $C$ is durably opaque. 
\end{lemma}



\section {A Modular Proof Technique}
\label{sec:modular-proofs}
In this section, we present a new approach to verifying durable opacity that allows one to leverage existing simulation-based proofs of opacity. An overview of the proof steps is shown in \reffig{fig:steps}. Given an existing opacity proof (step \circled{1}) that uses simulation against \TMS, the majority of the effort in the modularised proof method is the development of libraries \AM and \CM that handle memory operations and a proof of durable linearizability between the two. 
We exemplify this proof technique on a durable version of the STM algorithm \NOREC 
which we newly develop below. 



\begin{figure}[t]
    \centering
    \begin{tikzpicture}
    \node[draw](TMS2) {\footnotesize \TMS};
    \node[draw, below=1.8cm of TMS2] (IMPL) {{\footnotesize \IMPL}};

    \path (TMS2) -- node[rotate=-90]{$\ge$} 
                    node[thick,xshift=-10,circle,draw,left,inner sep=1pt, outer sep=1pt]{1}
                    node[thick,xshift=10,right,inner sep=1pt, outer sep=1pt]{(prove)}
           (IMPL);
    \node[draw, right = 1.9cm of TMS2] (DTMS2) {\footnotesize \DTMS};
    \node[draw, below = 1.8cm of DTMS2] (DIMPL) {\footnotesize \DIMPL};

    \path(DTMS2) -- node[rotate=-90] {$\geq$} 
                    node[thick,xshift=-10,circle,draw,left,inner sep=1pt, outer sep=1pt]{3}   
                    node[thick,xshift=10,right,inner sep=1pt, outer sep=1pt]{(check)}
(DIMPL);
    
    \draw [thick,-{>[scale=2.0]}] (IMPL) -- node[above, xshift=-0.1cm]{+ crash}
    node[yshift=-2,circle,draw,below,inner sep=1pt, outer sep=1pt]{2}
    (DIMPL);

    \node[draw, right = 2.5cm of DIMPL] (STMAM) {\footnotesize \DLIMPL{\AM}};

    \node[draw, above = 1.8cm of STMAM] (AM) {\footnotesize \AM};

    \draw [thick,-{>[scale=2.0]}] (DIMPL) -- node[above, xshift=-0.1cm]{modularise}
    node[yshift=-2,circle,draw,below,inner sep=1pt, outer sep=1pt]{4} (STMAM);

    \node[draw, right = 2.9cm of STMAM] (STMCM) {\footnotesize \DLIMPL{\CM}};
    \node[draw, above = 1.8cm of STMCM] (CM) {\footnotesize \CM};

    \path (AM) -- node{$\geq$ (prove)}     node[thick,yshift=-8,circle,draw,below,inner sep=1pt, outer sep=1pt]{5}
(CM);

    \draw [thick,-{>[scale=2.0]}] (STMAM) -- node[above]{replace}
    node[below]{\refthm{th:ref-context}}
(STMCM);

\draw[draw=black,dashed] ($(IMPL)+(-1,-0.8)$) rectangle ($(TMS2)+(1.1,0.8)$);
\draw[draw=black,dashed] ($(DIMPL)+(-1,-0.8)$) rectangle ($(AM)+(1.3,0.8)$);
\draw[draw=black,dashed] ($(STMCM)+(-1.3,-0.8)$) rectangle ($(CM)+(1.3,0.8)$);

    \node[above = 1.3cm of TMS2] (NVM_L) {Persistent};
    \node[above = 0.8cm of TMS2] (NVM_L) {memory only};
\node[above left = 1.3cm and -1.8cm of CM] (CM-L) {Persistent and volatile heap};
    \node[above left = 0.8cm and -1.8cm of CM] (CM-L) {Volatile implementation vars};

    \node[above right = 1.3cm and -1.5cm of DTMS2] (NVM_L') {Persistent heap };
    \node[above right = 0.8cm and -1.5cm of DTMS2] (NVM_L') {Volatile implementation vars};



    {\color{red} \bf draw boxes}
    \end{tikzpicture}
    \caption{Overview of proof steps}
    \label{fig:steps}
\end{figure}

\subsection*{Overview of the proof technique}
\label{sec:overview} 
The main idea behind the  proof steps is to gradually introduce more complex (fine-grained) interactions between persistent and volatile memory, as outlined in \reffig{fig:steps}. We start with models (the implementation of the STM called $\IMPL$ and $\TMS$) in which all reads and writes interact directly with persistent memory, thus crashes have no effect. 

Next, we introduce models $\DIMPL$ and $\DTMS$, where we assume memory is partitioned into two sorts: locations that the transactional memory implementations read from and write to (aka the {\em heap}), and variables that are used to implement the STM. E.g., for our running example (\NOREC in \reffig{fig:NOrec}), variables such as $glb$, $loc_t$ and $rdSet_t$ are implementation variables. In the models $\DIMPL$ and $\DTMS$, we assume that the operations on the heap are directly over persistent memory, whereas operations on implementation variables are over volatile memory. The former are assumed to be preserved upon a system crash, whereas the latter are lost. In our IOA models, the loss of implementation is modelled by setting the program counter of any running transactions to aborted when a crash occurs; since transactions are not restarted, this is equivalent to losing the local variables. Upon recovery, shared implementation variables must be reset since they are reused. For our example, the global counter $glb$ is reset to $0$ during recovery.

The next phase introduces $\DIMPL[\AM]$, which uses a library $\AM$ that manages reads and writes to the heap. This model is a simple refactoring of $\DIMPL$ and hence has the same memory model: all reads and writes performed by $\AM$ are directly over persistent memory, whereas those performed by $\DIMPL[\cdot]$ are on the transactional memory implementation and are hence volatile.  

The final phase refines $\AM$ into $\CM$ such that $\CM$ is durably linearizable w.r.t. $\AM$. Here, $\CM$ is a fine-grained implementation of $\AM$, and hence we assume that it operates over both volatile and persistent memory.

We now describe the main steps of our proof method, as outlined in \reffig{fig:steps} in more detail. 

\begin{itemize}[left=3mm] 
\item[\circled{1}] Our modular proof method starts with an existing
  simulation-based proof between an STM implementation, \IMPL, and
  TMS2, which establishes opacity of \IMPL. In our example case study,
  we consider $\NOREC$ as our implementation, which has already been proven correct with
  respect to $\TMS$~\cite{Lesani2012}. Our primary motivation for the new approach is to
  develop a durable version of an already opaque algorithm and avoid
  full re-verification of the durable version (against \DTMS).

\item[\circled{2}] To make sense of the adaptation of \IMPL to persistent memory,
  we start by assuming \IMPL runs directly on persistent memory, i.e.,
  all memory accesses are persistent, and there is no notion of
  volatile memory. Hence, there are no flush operations that transfer
  memory contents from volatile to persistent memory. Of course, \IMPL
  also does not contain a recovery operation since
  it has been designed to be opaque as opposed to durably opaque.

This step of the transformation therefore is to define a new version
of \IMPL, \DIMPL, that extends \IMPL with a ``crash-recovery''
operation. The purpose of this operation is to crash any live
transactions so that they are no longer able to execute and to
rollback the memory to a consistent state. We will require that the
crash-recovery introduced into \DIMPL is a refinement of the
$crashRecovery$ operation of \DTMS from \reffig{fig:tms2}. 

The introduction of a crash-recovery operation must be coupled with some small adjustments to the original algorithm. For instance, performing a write-back must be made atomic; a crash-recovery in the middle of a non-atomic write-back would leave the transactional memory heap in an inconsistent state. Details in the context of our running example are given in \refsec{sec:step2}. 



\item[\circled{3}] The next step is to verify that the transformed algorithm \DIMPL
  is durably opaque. To do this, we must adapt the existing simulation
  proof between \IMPL and \TMS to prove simulation between \DIMPL and
  \DTMS. Recall that the transformation from both \IMPL to \DIMPL and
  \TMS to \DTMS involves the introduction of a crash-recovery
  operation. Also recall that we assume both algorithms run directly
  on persistent memory, and no volatile memory is assumed in either
  case. Therefore, the effect of this crash-recovery operation in both
  cases is straightforward, and the adaptation of the proof is
  therefore straightforward too.

\item[\circled{4}] In the next step we adapt \DIMPL so that it relegates all memory
  operations to an external library.  We call this adapted algorithm
  \DLIMPL{\cdot}, which is \DIMPL but with calls to external operations
  that manage memory interactions. We must additionally develop a data
  type (library IOA), \AM, that handles these memory events.
  This enables one to define a composition \DLIMPL{\AM} (via a product
  of two IOA), where the library used by \DLIMPL{\cdot} is \AM.

There are two requirements for the library \AM. 
\begin{enumerate}
\item We must be able to show that the traces of \DLIMPL{\AM} are a
  subset of the traces of \DIMPL, i.e., \DLIMPL{\AM} is a {\em trace
    refinement} of \DIMPL. Note that by transitivity of refinement
  this means that the composition \DLIMPL{\AM} will be durably opaque.
\item We must allow \AM to be {\em implemented} by a concrete memory
  library that uses both a volatile and persistent memory so that the
  algorithm can ultimately be implemented in a non-volatile memory
  architecture.
\end{enumerate}

To satisfy both criteria while keeping the proof burden light, we make
\AM an {\em atomic} object, i.e., reads and write-backs are
coarse-grained atomic operations (see \reffig{fig:ADT}). Moreover,  \AM
operates directly on persistent memory, i.e., it does not use volatile
memory. With these restrictions in place, it becomes straightforward
to show trace refinement between the traces of \DIMPL and
\DLIMPL{\AM}.

\item[\circled{5}] In the last step, we develop \CM, a {\em durably linearizable}
  implementation of \AM, comprising a fine-grained concurrent memory
  library that operates over both volatile memory and persistent
  memory. The library \CM manages (persistent) logging to undo
  partially completed operations (in case of a crash and recovery) and
  flushing to ensure operations on volatile memory are made
  persistent.

More importantly (see \refthm{thm:combined} below), we obtain a trace refinement property: the traces of \DLIMPL{\CM}
projected onto the events of \DIMPL only (i.e., ignoring the
library calls) are a subset of the traces of \DLIMPL{\AM} projected
onto the events of \DIMPL.

\end{itemize}




\subsection{Step 1: \NOREC} 
\label{sec:step1}
We start by instantiating \IMPL to \NOREC~\cite{DBLP:conf/ppopp/DalessandroSS10}, which is given by the algorithm in Figure~\ref{fig:NOrec}.\footnote{In~\cite{Derrick+21}, we describe a procedure for transforming
pseudocode into IOA, which we use here.} \NOREC employs a deferred update
strategy: writes to shared state are first stored in a write
set and at commit time written to main memory -- if there are no
conflicts with other transactions. For conflict detection, \NOREC uses
value-based validation (see the operation \verb+TMValidate+).

To synchronise concurrent transactions, \NOREC uses a global counter {\tt
  glb} (initially $0$) and a local variable {\tt loc}, which is used
to store a copy of {\tt glb}. Each transaction maintains a local write
set, {\tt wrSet}, and a local read set, {\tt rdSet}. Inside its {\tt
  wrSet} a transaction records all the addresses that it attempts to
update and their values. The actual update of the memory takes place
inside the commit operation. An odd {\tt glb} indicates that a live
writing transaction attempts to commit. After a successful commit {\tt
  glb} is incremented so that its value is once again even. Thus a live transaction can determine whether another writing transaction has performed a commit operation by checking whether the value of {\tt glb} is equal to its local copy, {\tt loc}. Inside its {\tt
  rdSet}, a transaction records the addresses that it reads and their
corresponding values. Every time a transaction attempts to read
an address that is not inside its {\tt wrSet}, if {\tt loc} $\neq$
{\tt glb}, then the validation method is executed. The validation method
waits until the global lock is not held ({\tt glb} is even), then checks that the {\tt rdSet} is still valid (w.r.t., the current memory state). In the case of
writing transactions, the validity of {\tt rdSet} is also
checked at the commit stage. 

Operation \tmbegin\ copies the value of {\tt glb} into its local
variable {\tt loc} and checks whether {\tt glb} is even. If this is
so, the transaction is started. Otherwise, a writing transaction is in
progress, so the process attempts to start again by rereading {\tt
  loc}. The operation \verb+TMValidate+ checks if the transaction's
{\tt rdSet} is consistent with the current state of
memory
to {\tt glb}, and returns {\tt time}.

\tmread\ 
first checks if the transaction has already written the
address it attempts to read. In that case it
returns the address' value from its {\tt wrSet}.
Otherwise it
checks if {\tt loc} is up to date, i.e., equal to {\tt glb}. If it is
not up to date, another transaction has updated the
memory and {\tt rdSet} should be checked to ensure that its values are
consistent with the current state of the memory. The check is
performed by calling the operation \verb+TMValidate+.  If the {\tt
  rdSet} is found consistent, the {\tt loc} is updated with the value
of {\tt time} that \verb+TMValidate+ returns ({\tt R4}).The (address,
value) pair is then added in {\tt rdSet} for future validation ({\tt
  R6}). Finally, the value of the read address is returned. If the
{\tt rdSet} is not found consistent, the transaction aborts.

\tmwrite\ adds the (address, value) pair that is to be
written at the {\tt wrSet} of the transaction. The memory is updated
at the commit stage. The operation \verb+TMCommit+ first checks if the
transaction is a read-only transaction. If it is, then no further
checking is required and the transaction commits at {\tt E1}. If it is
not, {\tt E2} checks whether a concurrent writing transaction has been
committed. If no such commit has occured, the CAS at {\tt E2} succeeds, i.e., {\tt loc = glb}, so {\tt glb} becomes odd (meaning that the writing
transaction obtains the lock). If the CAS does not succeed a concurrent writing
transaction has been committed and {\tt rdSet} needs further
validation. {\tt E3} validates \verb+rdSet+ and updates the value of
{\tt loc}. By this, it prepares the transaction for another commit
attempt. When the commit has obtained the lock by making {\tt glb} odd, memory is updated with all the values from the write set in the loop at {\tt E4} and {\tt E5}. At {\tt E6} the
transaction releases the lock by making the {\tt glb} value even
again.
\begin{figure}[t]
\small
\noindent
\begin{minipage}[t]{0.45\columnwidth}
\begin{lstlisting}
Init:
I1 glb := 0;

$\tmbegin_t$:                         
B1 do $loc_{t}$ := glb; 
B2 until even($loc_t$)  
   return ok;
                                         
$\tmread_t$(addr):    
R1 if addr $\in$ dom($wrSet_t$) then 
     return $wrSet_t(addr)$;
R2 $v_t$ := *addr;
R3 if $loc_t$ $\neq$ glb then {
R4   $loc_t$ := TMValidate$_t$;
R5   goto R2;
   }
R6 $rdSet_t$.insert(addr,$v_t$); 
   return $v_t$;
   
   
$\tmwrite_t$(addr,val): 
W1 $wrSet_t$.insert(addr,val); 
   return ok;
   
\end{lstlisting} 
\end{minipage}
\hfill 
\begin{minipage}[t]{0.52\columnwidth}
\begin{lstlisting}[escapeinside={(*}{*)}]
$\tmend_t$:   
E1 if $wrSet_t$.isEmpty() 
     then return ok;
E2 while !cas(glb, $loc_t$, $loc_t$ + 1)    
E3   $loc_t$ := TMValidate$_t$;
E4 for  $\all$(addr,val) $\in$ $wrSet_t$ 
E5   *addr := val;
E6 glb := $loc_t$ + 2;  
   return ok; 
  
   
TMValidate$_t$:
V1 while true 
V2   $time_t$  := glb;
V3   if odd($time_t$) then goto V2;
V4   for $\all$(addr,val) $\in$  $rdSet_t$ do
V5     if *addr $\neq$ val 
          then abort;
V6   if $time_t$ = glb  
        then return $time_t$;

\end{lstlisting} 
\end{minipage}
\caption{The \NOREC algorithm. Line numbers for {\tt return}
  statements are omitted.}
\label{fig:NOrec}
\end{figure}

Correctness of \NOREC has been verified by Lesani et al.~\cite{Lesani2012} using the theorem prover PVS.
The proof proceeds via  showing a refinement relationship (simulation relation) between \NOREC and
\TMS~\cite{DGLM13}.  For the proof, \NOREC is first of all transformed into an IOA. 
Here, we only exemplify this on one operation, {\tt TMWrite} with one statement W1. The operation is split into three actions, 
an invocation and a response action (both external) plus a do action (internal).

\[ \action{\writeInv{t}{l, v}}
{pc_t = \pcReady}
{pc_t := \pcDoWrite(l, v)}
\action{{\tt DoWrite}_t}
{pc_t = \pcDoWrite(l,v)}
{pc_t := \pcWriteResp\\&
 wrSet_t := wrSet_t \oplus \{l \rightarrow v\}}
\action{\writeResp{t}}
{pc_t = \pcWriteResp}
{pc_t := \pcReady}     
 \] 

The proof of refinement between \NOREC and \TMS is carried out via the
construction of a sequence of IOA in between \TMS and \NOREC, and a
sequence of simulation proofs from one to the next IOA on this
sequence. We will not give all the details of this proof here, rather
concentrate on the key concepts and their relationships to the durable version.

\subsection{Step 2: Defining \NOREC with crash and recovery}
\label{sec:step2}


\begin{figure}[!t]
\small
\begin{minipage}[t]{0.45\columnwidth}
\begin{lstlisting}
$\tmread_t$(addr):    
R1 if addr $\in$ dom($wrSet_t$) then 
     return $wrSet_t(addr)$
R2 atomic {
     if owns = t $\lor$ owns = $\bot$
     then $v_t$ :=  *addr
     else $v_t$ := ? }
R3 if $loc_t$ $\neq$ glb then {
R4   $loc_t$ := TMValidate$_t$
R5   goto R2 
   }
R6 $rdSet_t$.insert(addr,$v_t$); 
   return $v_t$
\end{lstlisting}
\end{minipage}
\hfill 
\begin{minipage}[t]{0.54\columnwidth}
\begin{lstlisting}   
$\tmend_t$:   
E1 if $wrSet_t$.isEmpty() 
     then return ok;
E2 while !cas(glb, $loc_t$, $loc_t$ + 1)    
E3   $loc_t$ := TMValidate$_t$
E4 atomic { 
     if owns = $\bot$ then owns := t }
E5 atomic {
     for  $\all$ addr. (addr,val) $\in$ $wrSet_t$ 
        *addr := val }
E6 atomic { 
     if owns = t then owns := $\bot$ }
E7 glb := $loc_t$ + 2;  
   return ok; 

Recovery:   
RC1 glb := 0
RC2 owns := $\bot$ 
\end{lstlisting}
\end{minipage}

\caption{The {\tt TMRead}, {\tt TMCommit} and {\tt Recovery} operations of \DNOREC.
  Note that the loop at {\tt E5} now executes atomically.}
\label{fig:cnorec-comm}
\end{figure}

Next, we define an enhanced algorithm \DNOREC. \DNOREC plays the role of \DIMPL in \reffig{fig:steps}. 
It differs from \NOREC in three ways.

First, \DNOREC has an additional operation which 
abstractly models the occurrence of a crash and the subsequent recovery operation.\footnote{Note that crash and recovery could be modelled as two separate operations. However, we expect recovery to execute in our implementation immediately after a crash and before any new transactions are started, i.e., the crash and subsequent recovery are sequential. Thus, we simplify the model and combine the crash and recovery operations. } In particular, it (1) simulates crashes by ensuring that no transaction ``survives'' crashes, i.e., the currently running transactions cannot continue their operations, and (2) performs a recovery to bring the metadata (in \DNOREC, \texttt{glb}) back to the initial state.
As discussed above, in \DNOREC, we assume that the heap is persistent, thus the recovery part is almost empty. 
We give this additional operation directly as an IOA action: 

\[\action{crashRecovery}
{true}
{pc := \lambda t \in T.\ \sif\  pc_t \notin \{notStarted,committed\}\ \sthen\  aborted\ \selse\ pc_t \\
& glb := 0 \\
& owns := \bot 
}\] 
\noindent In \reffig{fig:cnorec-comm}, we present pseudocode
describing the \texttt{Recovery} procedure modelled by an atomic action.

Second, \DNOREC's commit operation is different to
that of \NOREC to deal with the fact that a crash can occur at
any time. \DNOREC must therefore update the shared memory with its write set {\em atomically}. In a later step (see \refsec{sec:cm-dlin}) we show how the write-back can be safely made non-atomic when using both volatile and non-volatile memory.

Third, to be compatible with the abstract library in step \circled{4}, we introduce an ownership variable, {\tt owns}, whose value is equal to a transaction iff that transaction currently has permission to write to the memory. In particular, ${\tt owns}$ is acquired by a transaction immediately prior to performing a write back ({\tt E4}), and released immediately after ({\tt E6}). A read from the (persistent) heap must return a random value in the presence of a concurrent writer since this indicates a potential data race between a reader and writer. In $\DNOREC$ (see \reffig{fig:cnorec-comm}), the read at line {\tt R2} reads from memory only if ${\tt owns} = t \lor {\tt owns} = \bot$ and otherwise returns a random value.
A read returning a random value in the presence of another writing transaction is unproblematic from the perspective of (durable) opacity since such read operations will either be revalidated, or if a revalidation is not possible, the reading transaction will abort. In particular, for the implementation to be (durably) opaque, a read must never return an illegitimate value even if it reads this value from memory.

The 
idea of using ownership in an interface to enforce atomicity has been
explored in prior work \cite{schellhorn2020adding}.
Variables akin to ownership are typically already present in correctness proofs of opacity since a transaction must have exclusive access to the shared memory during write back. For \NOREC the {\tt owns} variable is equivalent to the already existing auxiliary {\tt commitLock} variable in the proof by Lesani et al~\cite{Lesani2012}. 



\subsection{Step 3: Checking \DNOREC refines \DTMS}
We prove durable opacity of \DNOREC by showing that it refines \DTMS. This is straightforward to check for three reasons. (1) We make the write-back in {\tt TMCommit} of \NOREC atomic, and this trivially preserves behaviours of a non-atomic write-back. (2) The only new operation is {\it crashRecovery}, which preserves the original simulation relation used in the original proof by Lesani et al.  
(3) Reads from memory return an undefined value only when we know that the corresponding {\tt TMRead} operation will fail. 
\begin{lemma}
\label{lem:DTMS-BNOREC}
$\DNOREC \le \DTMS$. 
\end{lemma}
The proof has been mechanised in PVS~\cite{KIV-DNOREC}, and is an adaptation of the mechanised proof by Lesani et al.~\cite{Lesani2012} that shows that \NOREC refines \TMS. Their proof is structured into four layers as shown in \reffig{fig:lesani}. This structure keeps each refinement proof small, and design details of the \NOREC algorithm are incrementally introduced. The most abstract is the \TMS specification, which is shown to be refined by the next layer, {\sc NOrecAtomicCommitValidate}, where read validation and commit write back are atomic. The next layer, {\sc NOrecDerived}, introduces a fine-grained write back operation, but leaves the read validation atomic. Finally, \NOREC is shown to be a refinement of  {\sc NOrecDerived}, where reads and validation are split into separate atomic steps.  

The three changes needed between \NOREC and \DNOREC must be reflected in each of these layers as shown in \reffig{fig:lesani}, i.e., we obtain {\sc dNOrecAtomicCommitValidate} and {\sc dNOrecDerived}, which are analogues of {\sc NOrecAtomicCommitValidate} and {\sc NOrecDerived}, respectively. We have the following changes as highlighted in \reffig{fig:lesani}. 
\begin{itemize}
    \item {\sc dNOrecAtomicCommitValidate} is obtained from {\sc NOrecAtomicCommitValidate} by introducing a crash-recovery operation, which, like \DNOREC resets {\tt glb} to $0$. No other changes are necessary since transactional read and write operations are atomic.
    
    \item  {\sc dNOrecDerived} is derived from {\sc NOrecDerived} by introducing the crash-recovery operation described above, and additionally reintroducing an atomic write-back (since {\sc NOrecDerived} uses a fine-grained commit loop). From the perspective of the simulation proof, this introduces a minor change to the verification, whereby the linearization point is shifted. In {\sc NOrecDerived}, the line corresponding to the successful {\tt cas} at line {\tt E3} can be used as the linearization point since this is the point at which the commit lock (aka ownership) is taken. In the context of durable opacity, linearizing the commit at a successful {\tt cas} is no longer valid since the operation could still crash even after the {\tt cas} is successful. Thus, in the revised proof, we shift the linearization point to the atomic write-back itself. 

    \item  The differences between \NOREC and \DNOREC are already described above; the modified operations {\tt TMRead} and {\tt TMCommit} are shown in \reffig{fig:cnorec-comm}. Since this level splits the atomicity of {\tt TMRead}, in addition to the  changes described for {\sc dNOrecDerived}, we must allow reads to return a random value if the read is destined to fail (as discussed above). This allows one-one compatibility with the abstract library introduced in the next step. 
    Use of a read that returns a random value is unproblematic from the perspective of the proof since a key invariant for \NOREC is that no other transaction is performing its write back when a transaction is reading. 
\end{itemize}

\begin{figure}[t]
    \centering
    \begin{tikzpicture}
    \node[draw] (TMS2) {\footnotesize \TMS};
    \node[draw, below= of TMS2] (NACV) {{\footnotesize \sc NOrecAtomicCommitValidate}};

    \node[draw, below= of NACV] (ND) {{\footnotesize \sc NOrecDerived}};

    \node[draw, below= of ND] (NR) {{\footnotesize \sc NOrec}};

    \path (TMS2) -- node[rotate=90]{$\le$} 
          (NACV) -- node[rotate=90]{$\le$} 
          (ND) -- node[rotate=90]{$\le$} (NR);

    \node[draw, right= 8cm of TMS2] (DTMS2) {\footnotesize \DTMS};
    \node[draw, below= of DTMS2] (DNACV) {{\footnotesize \sc dNOrecAtomicCommitValidate}};

    \node[draw, below= of DNACV] (DND) {{\footnotesize \sc dNOrecDerived}};

    \node[draw, below= of DND] (DNR) {{\footnotesize \sc dNOrec}};


    \path (DTMS2) -- node[rotate=90]{$\le$} 
          (DNACV) -- node[rotate=90]{$\le$} 
          (DND) -- node[rotate=90]{$\le$} (DNR);

    \path[draw, dashed, thick, ->] (TMS2) --node[above]{\small+ crash, recovery} (DTMS2); 
    \path[draw, dashed, thick, ->] (NACV) --node[above]{\small+     \begin{tabular}{@{}l@{}} 
          crash, recovery
          \end{tabular}} (DNACV);
    \path[draw, dashed, thick, ->] (ND) --node[above]{\small
    + \begin{tabular}{@{}l@{}} 
        crash, recovery, \\ 
        atomic write-back
    \end{tabular}} (DND);
    \path[draw, dashed, thick, ->] (NR) --node[above]{\small
    + \begin{tabular}{@{}l@{}} 
        crash, recovery, atomic write-back, \\ 
         read ownership
    \end{tabular}} (DNR);
    \end{tikzpicture}
    \caption{Adapting Lesani et al's~\cite{Lesani2012} proof steps}
    \label{fig:lesani}
\end{figure}





\subsection{Step 4: Modularising \DNOREC and defining \AM} 
\label{sec:modularise}

\begin{figure}
\bgroup
\setlength{\tabcolsep}{1em}
\setlength{\myrowsep}{4em} 
\setlength{\myrowsepa}{3em} 
\setlength{\myrowsepb}{5.5em} 
\setlength{\myrowsepc}{6.5em} 
  \begin{tabular}{@{}l@{\qquad\quad}l@{}}

\action{\acquireob_t}
{true}
{{\bf if}\ owns = \bot\ {\bf then}\ owns := t}
&
\action{\OreadInv{t}{addr}} 
{lpc_t = \pcReady}
{lpc_t := \pcDoLibRead(addr)}
\\[\myrowsepa]
\action{\readob_t(addr;v)}
{true}
{{\bf if}\ owns = t \lor owns = \bot\\&
  {\bf then}\ v := mem(addr)\\&
  {\bf else}\ v :=\ ?
  }
&
\action{\doLibRead{t}{addr}}
{lpc_t = \pcDoLibRead(addr) 
}
{
  {\bf if}\ owns = \bot \lor owns = t\\&
  {\bf then}\  v := mem(addr)\\&
  {\bf else}\ v :=\ ? \\&
  lpc_t := \pcReadLibResp(v) }
\\[\myrowsepc]
\action{\releaseob_t}
{true}
{{\bf if}\ owns = t\ {\bf then}\ owns := \bot}
&
\action{\OreadResp{t}{v}}
{lpc_t = \pcReadLibResp(v)}
{lpc_t := \pcReady}
\\[\myrowsepa]
\action{\writeob_t(wrset)}
{owns = t}
{mem := mem \oplus wrset
  }
& \\ 
& \\
\action{\tt LibRecovery}
       {true}
{lpc := \lambda t: T. \\&
       \quad \textbf{if\ } lpc_t \neq \pcNotStarted\\& \quad \textbf{then\ } \pcCrashed \\&
       \quad \textbf{else\ } lpc_t 
}
& 
  \end{tabular}
\egroup
  \caption{Sequential Specification $\bbL$ (left) of the library
           and transitions for LibRead of the IOA \AM = $\duraut{\bbL}$ (right)}\label{fig:ADT}
\end{figure}

\begin{figure}[t]
\small
\noindent
\begin{minipage}[t]{0.45\columnwidth}
                                         
\begin{lstlisting}
$\tmbegin_t$:                         
B1 do $loc_{t}$ := glb; 
B2 until even($loc_t$)  
  return ok;

$\tmread_t$(addr):    
R1 if addr $\in$ dom($wrSet_t$) then 
     return $wrSet_t(addr)$
R2 $v_t$ := LibRead$_t$(addr)
R3 while $loc_t$ $\neq$ glb  
R4   $loc_t$ := TMValidate$_t$
R5   $v_t$ := LibRead$_t$(addr)
R6 $rdSet_t$.insert(addr,$v_t$); 
   return $v_t$

Recovery:
RC1 atomic { 
     LibRecovery;  
     glb := 0; }

$\tmwrite_t$(addr,val): 
W1 $wrSet_t$.insert(addr,val);  
   return ok;
\end{lstlisting} 
\end{minipage}
\hfill 
\begin{minipage}[t]{0.52\columnwidth}
\begin{lstlisting}[escapeinside={(*}{*)}]
$\tmend_t$:   
E1 if $wrSet_t$.isEmpty() 
      then return ok;
E2 while !cas(glb, $loc_t$, $loc_t$ + 1)    
E3   $loc_t$ := TMValidate$_t$
E4 $\acquireob_t$
E5 $\writeob_t$($wrSet_t$)
E6 $\releaseob_t$
E7 glb := $loc_t$ + 2;  
   return ok; 
  
   
TMValidate$_t$:
V1 while true 
V2   $time_t$  := glb
V3   if odd($time_t$) then goto V2
V4   for $\all$ (addr,val) $\in$  $rdSet_t$ do
V5     if $\readob_t$(addr) $\neq$ val 
       then abort
V6   if $time_t$ = glb  
        then return $time_t$

\end{lstlisting} 
\end{minipage}
\caption{The \DLNOREC{\cdot} algorithm with library calls that relegate memory operations to a library}
\label{fig:CNOrec}
\end{figure}

Next, we modularise \DNOREC by calling a library instead of directly
accessing shared memory. We start with a sequential library specification $\bbL$ (\reffig{fig:ADT}, left), which we convert into a concurrent durable IOA (\reffig{fig:ADT}, right) using the technique described in \refsec{sec:lin-dlin}. 

The modularised algorithm \DLNOREC{\AM} is given
in \reffig{fig:CNOrec}, 
where reads from memory occur through calls to the \readob\ operation, the write-back occurs via \writeob, and acquire/release of ownership via \acquireob\ and \releaseob, respectively. In the first instance, we start with an abstract library $\AM$ (see \reffig{fig:ADT}) that matches the code in \DNOREC exactly. Technically, moving atomic steps
of \DNOREC to library calls that execute
the same atomic step does not
change the algorithm. Its traces are unchanged
if the external invoke and response actions calling and returning from library
operations are hidden.

Note that it is crucial for the library to include operations for acquiring and releasing ownership. 
The specification $\bbL$ directly expresses
that concurrent \writeob\ calls are impossible since the precondition of \writeob\ requires that the calling thread is the current owner. 
This is exploited
when developing a concurrent implementation
such as \CM defined in the next subsection.
In particular, the correctness proof of \CM
does not have to prove linearizability
for two concurrent calls of \writeob,
which would have been necessary for
a library that only offers \readob\ and
\writeob\, without mentioning ownership.
Ownership therefore is used as a way to formalise
``linearizability under
constraints of not calling specific operations concurrently''
as ordinary linearizability (here: durable
linearizability).

Using the notation from \refsec{sec:ref-in-context}, \DLNOREC{\AM}
denotes the program in \reffig{fig:CNOrec} using the abstract memory
in \reffig{fig:ADT}.  The traces of   \DLIMPL{\AM}
include, as external actions, the external actions of both \DIMPL
and those of the library $\AM$. Let $traces(\DLIMPL{\AM})_{|\DIMPL}$
denote the traces restricted to just \DIMPL. The next lemma
establishes durable opacity of \DLNOREC{\AM} by stating that it
refines \DNOREC which we know to refine \DTMS (by
Lemma~\ref{lem:DTMS-BNOREC}) which itself is durably opaque (by
Theorem~\ref{theor:dtsm-sound}). 
\begin{lemma}
\label{lem:library}
$traces(\DLNOREC{\AM})_{|\DNOREC} = traces(\DNOREC)$. 
\end{lemma}

Note that we prove that the trace sets
are equal, not just subset. To do this
we prove that the preconditions of 
AM operations are always satisfied at their call sites in \mbox{$\DNOREC$[AM]}. 
A client that violates the precondition of an AM call at some call site would be deadlocked due to the semantics of IO Automata: a violated precondition of a transition means that it is disabled. The refinement would still be correct for
such a client, since IO automata refinement  as well as (durable) linearizability/opacity does not guarantee any liveness. In the extreme an empty implementation that has no transitions enabled at all is correct, though not useful.
Proving that preconditions of calls hold,
together with the fact that 
a sequential program, when translated 
to an IO automaton always has its next step enabled, guarantees that deadlocks are avoided in our case study.

\subsection{Step 5: Defining \CM and proving durable linearizability}
\label{sec:cm-dlin}

So far, the read/write operations on transactional variables have existed entirely on persistent memory. Our final task, therefore, is to develop a concrete library, \CM, that is durably linearizable w.r.t. \AM and manages low-level read/write operations across volatile and persistent memory (see \reffig{fig:steps}). The following theorem ensures that it is safe to perform such a replacement without violating durable opacity.

\begin{theorem} 
\label{thm:combined}
If (1) \CM is durably linearizable w.r.t.~$\bbL$, (2) \AM equals $\duraut{\bbL}$ and (3) \DLIMPL{\AM} 
  is a refinement of \DTMS,
  then \DLIMPL{\CM} is durably opaque.
\end{theorem}
\begin{proof}
The proof is by applying \refthm{th:ref-context}. Durable linearizability
  of \CM to $\bbL$ is equivalent to \CM being a refinement of the
  canonical IOA, \AM, which has been shown
  in Lemma~\ref{refines-CM-AM}.
  \DLIMPL{\CM} and \DLIMPL{\AM} can be constructed
  as the product IOA of \DLIMPL{\cdot} and \CM/\AM, respectively. 
  The shared external
  actions and steps between both IOA are the invocations and
  responses of library operations, together with the crash. We assume that the $run_t$ action of \AM and
  \CM is synchronised with the $\beginInv{t}$
  action of the \DLNOREC{\cdot} which starts a transaction.
  The theorem states that \DLIMPL{\CM} is a refinement of \DLIMPL{\AM}, implying that $traces(\DLNOREC{\CM})_{|\DNOREC} \subseteq traces(\DLNOREC{\AM})_{|\DNOREC}$ hiding invocations and responses of library operations. 
  Since \DLIMPL{\AM} refines \DTMS (Lemmas~\ref{lem:DTMS-BNOREC} and~\ref{lem:library}), and by transitivity of refinement we get that
  \DLIMPL{\CM} refines \DTMS, implying durable opacity.
\end{proof}

\medskip
We now describe the instance of \CM that we use (see \reffig{fig:CDT}).  \CM implements \AM on an architecture with persistent and volatile memory: 
 instead of writing directly to persistent memory $mem$ (as in the case of \AM), \CM first writes to the concrete volatile memory, $vmem$, and this is later flushed to the concrete persistent memory, $pmem$. 


Functional correctness is not affected when all transactions read and write to $vmem$ instead of $mem$. However, after a crash 
the data in $vmem$ is lost, and computation resumes from the state of $pmem$.
Therefore, to ensure durable opacity, we have to ensure that $pmem$ is updated during a commit so that the memory snapshot that results
from the successful commit is available even after a crash. For a lazy STM implementation like \NOREC, committing the
write set is the only place in the code which writes to memory,
so the implementation must update both $vmem$ and $pmem$ during a commit write back.

In \NOREC, a crash occurring partway through a commit write back may result in an inconsistent memory state, i.e., one that is not a snapshot
of the successfully completed transactions.
We treat transactions that crash during (or before) a commit write back to be an aborted transaction, thus any memory updates performed by a partially completed write back operation
must be reverted. To make this possible, we keep a {\em persistent}
log $plog$ that stores old values for those locations of the write
set that have already been committed. 

This leads to the following algorithm
for committing the given write set $wrSet$:
\begin{lstlisting}[escapeinside={(*}{*)}]     
     for $\forall$ (addr, val) $\in$ $wrSet$ do
          oldv := *addr;
          plog := plog $\oplus$ {addr $\mapsto$ oldv};
          *addr := val;
          flush(addr);
     plog := $\emptyset$;
\end{lstlisting}
\noindent
In KIV, the abstract code 
``\texttt{for} $\forall$ {\tt (addr, val)} $\in wrSet$''
is realised as a while loop, that iterates over 
the write set.  Translating to steps 
of an IOA, this gives the steps
shown in \reffig{fig:CDT}. The first action $W1$
is the loop test, 
that checks whether $wrSet$ is empty.
In case it is not, an $addr$ is chosen in step $W2$,
and the four instructions of the loop body above are executed as steps $W3$ to $W6$. Flushing moves $vmem(addr)$ to $pmem(addr)$. $addr$ is 
then removed from $wrSet$ in step $W7$ which jumps
back to the loop test. When $wrSet$ is empty, the loop is left and step $W8$ resets
the persistent log. In addition to the program steps the IO automaton for \CM includes a ${\tt flush}(l)$ step (with an internal action) that models flushing a memory location $l$ that is possible at any time.

On a crash, the log is used
to undo the partial commit\footnote{In the IOA, the recovery executes $vmem := pmem \oplus plog$; the KIV specification uses a recovery program that writes each log entry separately in a loop.}. When the write set has been
fully committed, the log is cleared, and clearing the log at $W8$
becomes the linearization point of the implementation of commit.
After this point the transaction has successfully committed.
\begin{figure}[hbt!]
{\small
\bgroup
\begin{tabular}{@{}l@{\quad\quad}l@{}}
\action{\OwriteInv{t}{wrSet}} 
{lpc_t = \pcReady \land owns = t}
{lpc_t := \pcwone(wrSet)}
&
\action{\wone{t}}
{lpc_t = \pcwone(wrSet)}
{lpc_t := {\bf if}\ (wrSet \neq \, \emptyset)\\&
\phantom{lpc_t :=\ }{\bf then}\  \pcwtwo(wrSet)\\&
\phantom{lpc_t :=\ }{\bf else}\ \pcweight}
\\[\myrowsepa] 
\\
\action{\wtwo{t}}
{lpc_t = \pcwtwo(wrSet) \\&
 SOME\ addr.\ addr \in dom(wrSet)
}
{lpc_t := \pcwthree(wrSet, addr)}
&
\action{\wthree{t}}
{lpc_t = \pcwthree(wrSet, addr) }
{lpc_t := \pcwfour(wrSet, addr)\\&
  oldv_t := vmem(addr)}
\\[\myrowsepa] 
\\
\action{\wfour{t}}
{lpc_t = \pcwfour(wrSet,addr)}
{lpc_t := \pcwfive(wrSet,addr) \\& 
plog := plog \oplus \{addr \to oldv_t\}}
&
\action{\wfive{t}}
{lpc_t = \pcwfive(wrSet,addr) }
{lpc_t := \pcwsix(wrSet,addr) \\&
  vmem(addr) := wrSet(addr)}
\\[\myrowsepa] 
\\
\action{\wsix{t}}
{lpc_t = \pcwsix(wrSet,addr)}
{lpc_t := \pcwseven(wrSet,addr)\\&
pmem(addr) := vmem(addr)}
&
\action{\wseven{t}}
{lpc_t = \pcwseven(wrSet, addr) }
{lpc_t := \pcwone(wrSet \backslash \{(addr, wrSet(addr))\}) 
}
\\[\myrowsepa] 
\\
\action{\weight{t}}
{lpc_t = \pcweight }
{
lpc_t := \pcWriteLibResp\\&
 plog := \emptyset}
&
\action{\OwriteResp{t}{}} 
{lpc_t := \pcWriteLibResp}
{lpc_t := \pcReady}       
\\[\myrowsepb]

\action{\rone{t}}
{
lpc_t = \pcDoLibRead(addr)
}
{
v := vmem(addr)\\ &
lpc_t := \pcReadLibResp(v)
}
& 
\action{run_{t}}
{lpc_t = \pcNotStarted}
{lpc_t := \pcReady}
\\[5em]

\action{\tt LibRecovery}
  {true}
  {lpc := \lambda t: \Tids.\ \sif\  lpc_t \neq ready\\
   & \quad\quad\ \  \sthen\ crashed \ \selse\ lpc_t \\
  & owns := \bot \\
  & vmem := pmem \oplus plog \\
  & pmem := pmem \oplus plog\\
  & plog := \emptyset 
  }
  &
  \action{{\tt flush}(l)}
         {true}
         {pmem(l) := vmem(l)}
\end{tabular}

\egroup
\smallskip

}
\caption{Transition relation
  of \CM. Transitions for Acquire/Release, as well as invoke and response transitions for read are the same as in \AM.} \label{fig:CDT}
\end{figure}

For the concrete library $\CM$ we have shown the following result. 

\begin{lemma}\label{refines-CM-AM}
\CM refines \AM. 
\end{lemma}

Since \AM is $\duraut{\bbL}$ for the sequential specification $\bbL$ of the library given in \reffig{fig:ADT}, we get the following corollary of 
Lemma~\ref{durlin-equiv-refine-canonical}.

\begin{corollary}
\CM is durably linearizable to $\bbL$. 
\end{corollary}

Lemma~\ref{refines-CM-AM} has been mechanically proven in the theorem prover KIV~\cite{KIV-Haehnle2022}. Both the IOA for \AM and for \CM are specified in KIV
by giving labelled programs which generate a predicate logic
specification of the transition relation. Specifications
and proofs are online at \cite{KIV-DNOREC}.

The refinement from \AM to \CM is proven in two steps.
First an invariant for \CM is proven which 
overapproximates the set of reachable states. The invariant is then
used in place of $reach(C)$ in the proof of a forward simulation
according to Def.~\ref{def:for-sim}.

The invariant of \CM consists of a global invariant and
local assertions. The global invariant
\[
  {\bf if}\ owns = \bot\ 
  {\bf then}\ vmem = pmem \land plog = \emptyset\ {\bf else}\ vmem \oplus plog = pmem \oplus plog
\]
\noindent
states that as long as there is no writer that commits a write set,
volatile and persistent memory agree, and the log is empty.
Otherwise overriding the volatile and
persistent memory with the log gives the same result: both
result in the memory snapshot at the start of the commit.

The local assertions give formulas that hold 
when a transaction $t$ is at a specific program counter, $lpc_t$. (We use $lpc$ here to distinguish the library program counter.)
As an example, while $t$  is executing the write operation
($lpc_t$ is one of $W1$ to $W8$) it has write ownership
($owns = t$). The KIV specification specifies
this implication (and many more) as pairs of a label range and a formula.
The full invariant is generated as a conjunction of all 
local assertions, that is universally quantified over all $t$, together with the global invariant. To have
a thread-modular proof of the assertions,
a rely predicate $rely(t, s, s')$ is specified that the 
steps of all other transactions $t' \neq t$ (from state $s$ to $s'$) and flush steps of the system must
satisfy. Assertions for thread $t$
and the global invariant are shown to be stable
with respect to this predicate.
In our case the rely predicate consists of three formulas.
\[
owns = t \rightarrow vmem = vmem' \land owns = owns' \land plog = plog'\\
owns = t \rightarrow between(pmem, pmem', vmem)\\
owns \neq t \rightarrow owns' \neq t
\]
\noindent
The first ensures that while a transaction $t$ is writing
other transactions will leave $vmem$, $owns$ and $plog$ unchanged.
The second asserts that while
a writer is running, $pmem$ may only be changed by system flushes: $between(pmem, pmem', vmem)$ asserts that all values $pmem'(l)$ will either still be $pmem(l)$ or be the flushed value $vmem(l)$. 
The third guarantees that steps of other threads
cannot make $t$ the writer.

The specification of \CM in KIV also fixes the linearization
points of \CM, by defining non-$\tau$ actions for such steps.
Reading linearizes at $\pcDoLibRead$, when the value is read from
volatile memory. Committing linearizes at $W8$, when
the log is set to empty.

The forward simulation between \CM and \AM consists
of a global part and a local part for every transaction $t$ too.
The global part simply states that $owns$ of \AM and \CM
are identical and that the abstract memory $mem$
of \AM is always equal to $vmem \oplus plog$ (the invariant
implies that it is then equal to  $pmem \oplus plog$ as well). 

The local part of the simulation for thread $t$
gives a mapping between program counter values of \CM
and \AM in the obvious way. As an example, since
the linearization point of commit is at $W6$, all
$lpc_t$ values before and including $W6$
are mapped to $\pcDoLibWrite(wrSet_t)$,
while $lpc_t = \pcWriteLibResp$  is mapped to $\pcWriteLibResp$.
The local part of the simulation also ensures that
input received by an operation of \CM that is stored in
a local variable is equal to the corresponding input of
\AM, and similar for the outputs.

With this forward simulation, the proof has to
show a commutativity for every step of \CM
according to Definition~\ref{def:for-sim}.
The proofs of this refinement in KIV are simple.
Three days of work were required to set up the specifications and to
do the proofs, which have ca.~300 interactive steps.

The final step is to combine the steps above instantiating \refthm{thm:combined}, resulting in the corollary below. 

\begin{corollary}
\label{cor:combined}
If (1) \CM is durably linearizable w.r.t.~$\bbL$, (2) \AM equals $\duraut{\bbL}$ and (3) \DLNOREC{\AM} 
  is a refinement of \DTMS,
  then \DLNOREC{\CM} is durably opaque.
\end{corollary}

Note, that in the implementation 
\CM the $owns$ variable
is an auxiliary variable, that has no
effect on computations.
Therefore the final program code 
equivalent to the IOA \DLIMPL{\CM}
can omit the variable together with the calls
to {\tt LibAcquire} and {\tt LibRelease}.

\section {Related Work}
\label{sec:related-work}


The literature around persistent memory has grown remarkably quickly. Below we provide a snapshot of some related work, focussing in particular on correctness and atomicity. 

\subsection{Correctness conditions}
Constructing robust shared objects for NVM requires the development
of criteria that provide meaningful guarantees in the presence of
crashes.  Linearizability \cite{HeWi90}, is one of the most
well-known, broadly used, correctness conditions for concurrent objects.
Several correctness conditions attempt to adapt linearizability to
histories that include crash events. 

As mentioned before, durable
linearizability \cite{DBLP:conf/wdag/IzraelevitzMS16} extends the
events that can appear in an abstract concurrent history with crash
events. Crashes are considered global events. Durable linearizability
expects that no thread survives after a crash, thus a thread can
operate only in one crash-free region. On the contrary, strict
linearizability \cite{aguilera2003strict}, consider crashes to be
local to the threads that they occur. Under this condition, operations
that are not subjected to a failure can take effect between their
invocation and response.  In the case that a thread crashes while
executing an operation, it requires this operation to take effect
between its invocation and the crash, but not after the
crash. Operations that are disrupted by a crash either take effect or abort when a crash occurs. 
Guerraoui and Levy~\cite{guerraoui2004robust} have defined two more
correctness conditions that extend linearizability, {\em persistent
atomicity}, and {\em transient atomicity}. Persistent atomicity requires that,
in the event of a crash, every pending operation on the crashed thread either takes effect or aborts before a subsequent operation of the same
thread is invoked, noting that an operation may take effect after a crash. Transient atomicity relaxes this condition
further, by allowing an incomplete operation to take
effect before a subsequent write response of the same thread.  
Berryhill et
al.~\cite{berryhill2016robust} have proposed {\em recoverable
linearizability}, which requires every pending operation on a thread to take effect or abort before the thread linearizes another operation. This condition does not 
provide consistency around the crash --- a thread can perform an
operation on some other object before coming back to the pending
operation causing ``program order inversion''. 
The main disadvantage of strict linearizability, persistent atomicity and transient atomicity is that they are not compositional. On the other hand, durable and recoverable linearizability are compositional. 

Several models, both hardware and software
specific, aim to define the correctness of the order in which writes
are persisted in NVM.  Pelley et al.~\cite{pelley2014memory} described
various such low-level models including {\em strict persistency} and relaxed
persistency models such as {\em epoch persistency} and {\em strand
persistency}. Those models consider hardware to be able to track
persist dependences and perform flushes in a manner described by the persistency model. Izraelevitz et
al. \cite{DBLP:conf/wdag/IzraelevitzMS16} gave formal semantics to
epoch persistency which corresponds to real-world explicit ISAs, where
flushes are issued explicitly with dedicated instructions by the
respective application. Raad et al. \cite{raad2019weak} developed
declarative semantics that formalise the persistency semantics of
ARMv8 architecture. \cite{raad2019persistency} propose persistency semantics for the Intel-x86 Architecture and \cite{cho2021revamping} provides view-based and axiomatic persistency models for Intel-x86 and ARMv8. 

On the language level, Kolli et
al.~\cite{kolli2017language} proposed an acquire and release
persistency model based on the acquire-release consistency of C++11. Furthermore,  Raad et al. \cite{raad2020persistent} and Bila et al.~\cite{DBLP:conf/esop/BilaDLRW22} have developed program logics for reasoning about persistent programs on Intel-x86, based on the Owicki-Gries proof system.

Regarding transactional memory \cite{harris2010transactional}, not
many correctness conditions have been adapted to the persistent memory
setting.  Raad et al~\cite{raad2019weak} base their framework for
formalising ARMv8 to a persistent variant of serializability (PSER)
under relaxed memory. Even though serializability provides simple
intuitive semantics, it does not handle aborted transactions.
TimeStone \cite{krishnan2020durable} and Pisces \cite{gu2019pisces}
are recent persistent transactional memories that guarantee snapshot
isolation \cite{berenson1995critique}, which is weaker than serializability, and hence opacity.

\subsection{Persistent Transactional Memory (PTM)}

Mnemosyne \cite{volos2011mnemosyne} provides a low-level interface to persistent memory with high-level transactions based on TinySTM \cite{felber2008dynamic}. The Mnemosyne transaction system combines lazy version management with the eager conflict detection that encounter-time locking provides. The lazy version management is implemented with a redo log, which has been chosen to reduce ordering constraints. The new writes to persistent memory are kept in a redo log and are buffered in the volatile memory. When a write transaction commits, it flushes the log to the persistent memory and optionally writes back the new values. Unlike TMs that use undo logging, the write transactions do not update the memory until they commit.  This adds an overhead to read transactions, since they should recognise the modified values, but not yet committed values, and then return them from the buffer. Moreover, the size of the log increases proportionally to the size of the transaction, potentially making commits time consuming.
Mnemosyne uses a global array of volatile locks to implement encounter-time locking. Every memory location is associated with a lock. Prior to accessing a memory location, the transaction identifies its associated lock and tries to acquire it. In the case that the operation succeeds, it adds the lock to the lock-set. Otherwise, it aborts and releases all the locks contained in its lock-set.

NV-heaps \cite{coburn2011nv} is a persistent object system that aims to integrate persistent objects into conventional programs, and furthermore seeks to prevent safety bugs that occur in predominantly persistent memory models, such as multiple frees, pointer errors etc. NV-heaps only handle updates to persistent memory inside transactions and critical sections. It uses ACID transactions to guarantee the consistency of persistent objects in the face of system failures. Specifically, NV-heaps rely on atomic sections that log all the updates of the non-volatile memory to provide fine-grain consistency. Each transaction keeps a volatile read log and a non-volatile write log.  NV-heaps provide eager conflict detection for writes. The system keeps a copy of the objects that are going to be modified by write transactions in an undo log. In this way, the modifications to an object can be rolled back in the case of an atomic section abort or a system failure. Each log update needs an epoch barrier, which affects the overall performance.
Before a transaction tries to modify an object, its atomic section attempts to take ownership of the object by acquiring a volatile lock in a table of ownership records. In case of success, the entire object is copied into the write undo log and the transaction proceeds to modify the object. Otherwise, the atomic section retries. To read an object, NV-heaps store a pointer to the object and its current version number in the read log. The version numbers help in detecting read conflicts at access time.

Unlike persistent transactional memories that provide durable transactions via undo and redo logs, Romulus \cite{correia2018romulus} provides durable transactions by keeping two copies (main and back) of the data in non-volatile memory and ensuring that at any time at least one of the copies is consistent. When a transaction begins, any modification of the data that is caused by the user-code is immediately flushed to the main copy.  Before a transaction ends, the modifications in the main copy are copied to the back copy.  If a failure occurs when the copying from main to back takes place, then the recovery procedure copies the contents of main to back. In the same way, if a failure occurs while the modification of main is taking place, the recovery procedure copies the contents of back to main. In order to avoid full replication of the data of the main to the back, Romulus introduces a volatile redo log that tracks the addresses of the modified data. At the end of the transaction, only those addresses are flushed to the back.  There are two implementations of Romulus available, one with a scalable reader-writer lock and another that uses a universal construct and supports wait-free read-only transactions.

OneFile \cite{ramalhete2019onefile} is a wait-free PTM that supports durably linearizable transactions. It uses a redo log for durability and a time based concurrency control. In this design, every thread maintains a redo log as write set that can be read by other threads in order to help the completion of the ongoing transaction, but does not maintain a read set. All writing transactions are associated with a unique sequence number that allows their serialization.  A technique that is similar to TL2 \cite{ShavTL2} and also uses sequence numbers is applied to ensure consistency of read operations.  The design of OneFile allows write transactions to run concurrently with read-only transactions. There are two variants of the OneFile available: one with lock-free progress and providing bounded wait-freedom.

QSTM \cite{beadle2020nonblocking} is a non-blocking persistent transactional memory. Its design is based on RingSTM \cite{DBLP:conf/spaa/SpearMP08} enhanced with a redo log based on the persistent lock-free queue of Friedman et al. \cite{DBLP:conf/ppopp/FriedmanHMP18}. Each transaction maintains a read and a write filter. The entries of the redo log represent the live transactions. Each entry  consists of a pointer to the transaction’s write set (this allows any thread to perform the writes of a committed transaction), a unique timestamp associated with the represented transaction and its write filter.  The validation mechanism is taking place within the read operation. Each transaction while reading is checking if its read filter conflicts with any write filter of the committed transactions. If so, it aborts.  Queue entries are deleted only when their respective writes are persisted. Beadle et al \cite{beadle2020nonblocking}  provide several correctness arguments of QSTM.  Specifically, they argue that QSTM is linearizable as a single concurrent object, durably linearizable,  and lock-free.
Compared to OneFile, QSTM uses significantly less space due to the fact that it does not require modifications in data declaration or the use of {\tt cas} and {\tt LL}/{\tt SC} instructions. However OneFile achieves higher throughput than QSTM, due to QSTM’s global log.




\subsection{Generic Approaches to Persistency}

Apart from PTMs, several generic frameworks have been developed to tackle the
problem of consistency under persistent memory.  Indicatively, Naama Ben-David et al. \cite{ben2019delay}
developed a system that can transform programs that consist of read,
write and {\tt cas} operations in shared memory, to persistent memory. The system
aims to create concurrent algorithms that guarantee consistency after
a fault. This is done by introducing persist checkpoints, which record
the current state of the execution and from which the execution can
continue after a fault. 

Izraelevitz et al.~\cite{izraelevitz2016failure} develop and implement
a logging mechanism based on undo and redo log properties named JUSTDO
logging and introduce the concept of FASE (failure-atomic
sections). This mechanism aims to reduce the memory size of log
entries while preserving data integrity after crash occurrences.
Unlike optimistic transactions \cite{chakrabarti2014atlas}, JUSTDO
logging resumes the execution of interrupted FASEs at their last store
instruction, and then executes them until completion. One disadvantage
of this strategy is that the FASEs cannot be rolled back after a
system failure. As a consequence, there is no tolerance of bugs inside
the FASEs. In this system, it is assumed that the cache memory is
persistent, and the system also requires that all load/store
instructions access persistent data. A small log is maintained for
each thread, that records its most recent store within a FASE. The
small per thread logs simplify the log management and reduce the
memory requirements.


\section{Conclusion}

In this paper, we use {\em durable opacity} as a correctness condition for STMs running on non-volatile hardware architectures. We have proposed an abstract specification 
\DTMS which is durably opaque and have shown how this can be employed in refinement-based proofs of 
durable opacity. We have furthermore developed a {\em modular} proof technique for such refinement proofs, 
separating out the proof of durability from that of opacity. We have exemplified this proof technique 
on the STM \NOREC for which we have -- to this end -- developed a version adequate for non-volatile memory. 


Our proof technique is inspired by work on the verification of a Flash file system by Schellhorn et al.~\cite{schellhorn2020adding}. 
Although this prior work does not target NVM or STMs, it also uses ownership as a mechanism for restricting 
concurrency at the interface of a library. The development,
which refines an abstract POSIX-compatible file system specification 
in several steps to the Linux interface  MTD for flash hardware,
 uses intermediate  layers (similar to the library
\AM used here) to introduce caches for flash pages \cite{DBLP:conf/ifm/PfahlerEBSR17}
and for file-content \cite{Bodenmueller-fac2022}, though the correctness
criteria defined in these papers are strictly weaker than durable linearizability.

Our new proof technique (outlined in \reffig{fig:steps}) describes a
similar technique in the setting of STMs. In particular, we provide an
abstract library that operates directly on persistent memory and a
concrete implementation that uses both volatile and persistent
memory. The original STM is placed in an execution context that could
have system crashes and uses the abstract and concrete library to
perform memory operations. Given an STM that already refines \TMS (and
hence is opaque), the bulk of the verification effort using our method
is focused on verifying durable linearizability between the abstract
and concrete libraries.
This proof is the only one that has to consider the distinction between volatile and persistent memory. In our case it is not very difficult when the correct ownership annotations are used.
In contrast, a non-modular proof would have to re-do the already complex opacity proof in the more complex setting where the distinction between volatile and persistent memory is present.
\paragraph{\em Future work.} We conjecture that the libraries
\AM and \CM that we have defined
could be used to transform other opaque algorithms into durably opaque
algorithms when the STMs use a lazy write-back commit with mutual
exclusion between the write-back operations. Other types of STMs,
e.g., TL2~\cite{ShavTL2} (which uses a per-location lock to allow
concurrent write-backs) and TML~\cite{DalessandroDSSS10} (which uses
an eager write-back mechanism) cannot use the libraries \AM and \CM
that we have developed directly. Whether the modular library based
approach presented in this paper applies to these other algorithms as
well remains a future research topic.

The memory model that we have assumed is strong, only making a distinction between volatile and persistent memory. The  writes themselves are assumed to be sequentially consistent, and no intra-thread reordering is possible. In reality, programs executed in platforms such as persistent x86-TSO~\cite{raad2020persistent,cho2021revamping}, in which instructions may be reordered due to the effects of both persistency and Total Store Order (TSO). In future work, we aim to extend our methods to additionally take such effects into account. In particular, it would be interesting to know whether \CM could be further refined and integrated with the remainder of the system in a modular manner. For the persistent TSO model, such proofs could build on  existing logics, e.g., \cite{raad2020persistent,DBLP:conf/esop/BilaDLRW22}, but may require new theories for refinement.  


\bibliographystyle{alphaurl}
\bibliography{references,references2}










   






 


\end{document}